\documentclass[a4paper,11pt]{article}
\usepackage{jheppub} 
\usepackage{lineno}
\usepackage{graphicx}
\usepackage{amsbsy,amsmath}
\usepackage{dsfont}
\usepackage{amsfonts}
\usepackage{bbold}
\usepackage{filecontents}
\usepackage{mathtools}
\usepackage{empheq}
\usepackage{slashed}
\usepackage{microtype}
\usepackage[macrosonly]{chet}
\usepackage{xcolor}
\definecolor{darkblue}{rgb}{0.1,0.1,.7}
\definecolor{darkgreen}{rgb}{0,.5,0}
\usepackage{hyperref}
\usepackage{booktabs}
\usepackage{array}
\usepackage{mdframed} 
\newmdenv[linecolor=black,linewidth=1pt,roundcorner=6pt, backgroundcolor=gray!10,skipabove=10pt,skipbelow=10pt,innerleftmargin=10pt,innerrightmargin=10pt, innertopmargin=8pt, innerbottommargin=8pt]{mybox}

\newcommand{\cO}{{\cal O}}

\DeclareMathOperator{\Tr}{Tr}

\renewcommand{\arraystretch}{0.8}

\renewcommand{\a}{\alpha}
\renewcommand{\b}{\beta}

\newcommand{\m}{\mu}

\newcommand{\bphi}{\bar{\phi}}

\def\be{\begin{equation}}
\def\ee{\end{equation}}
\def\bea{\begin{eqnarray}}
\def\eea{\end{eqnarray}}
\def\ba{\begin{array}}
\def\ea{\end{array}}
\def\bi{\begin{itemize}}
\def\ei{\end{itemize}}

\def\Tr{{\rm Tr}}

\newcommand{\beq}{\begin{equation}}
\newcommand{\eeq}{\end{equation}}
\newcommand{\beqn}{\begin{eqnarray}}
\newcommand{\eeqn}{\end{eqnarray}}
\newcommand{\bga}{\begin{align}}

\def\dalemb#1#2{{\vbox{\hrule height .#2pt
\hbox{\vrule width.#2pt height#1pt \kern#1pt
\vrule width.#2pt}
\hrule height.#2pt}}}

\def\a{\alpha}
\def\b{\beta}

\def\z{\zeta}

\def\bz{{\overline{z}}}

\def\sc{{\hat{c}}}

\def\cI{{\cal I}}
\def\cZ{{\cal Z}}
\def\ze{\zeta}
\def\bze{\bar{\zeta}}

 \def\IZ{\relax\ifmmode\mathchoice
 {\hbox{\cmss Z\kern-.4em Z}}{\hbox{\cmss Z\kern-.4em Z}}
 {\lower.9pt\hbox{\cmsss Z\kern-.4em Z}}
 {\lower1.2pt\hbox{\cmsss Z\kern-.4em Z}}\else{\cmss Z\kern-.4em Z}\fi}
 \def\IB{\relax{\rm I\kern-.18em B}}
 \def\IC{{\relax\hbox{$\inbar\kern-.3em{\rm C}$}}}
 \def\Ic{{\relax\hbox{$\inbar\kern-.22em{\rm c}$}}}
 \def\ID{\relax{\rm I\kern-.18em D}}
 \def\IE{\relax{\rm I\kern-.18em E}}
 \def\IF{\relax{\rm I\kern-.18em F}}
 \def\IG{\relax\hbox{$\inbar\kern-.3em{\rm G}$}}
 \def\IGa{\relax\hbox{${\rm I}\kern-.18em\Gamma$}}
 \def\IH{\relax{\rm I\kern-.18em H}}
 \def\II{\relax{\rm I\kern-.18em I}}
 \def\IK{\relax{\rm I\kern-.18em K}}
 \def\IP{\relax{\rm I\kern-.18em P}}

\def\Tr{{\rm Tr}}
 \font\cmss=cmss10 \font\cmsss=cmss10 at 7pt
 \def\IR{\relax{\rm I\kern-.18em R}}

\def\cO{{\cal O}}

\def\a{\alpha}
\def\b{\beta}

\def\m{\mu}

\def\cQ{{\cal Q}}
\def\cD{{\cal D}}

\def\cF{{\cal F}}

\title{\boldmath The thermal representation of conformal ladder integrals}

\author{Manthos Karydas$^a$, Songyuan Li$^b$, Anastasios C. Petkou$^b$ and Matthieu Vilatte$^c$}
\affiliation{$^a$ Physics Division, Lawrence Berkeley National Laboratory, Berkeley, CA 94720, USA}
\affiliation{$^b$ Laboratory of Theoretical Physics, School of Physics,
 Aristotle University of Thessaloniki, 54124 Thessaloniki, Greece}
\affiliation{$^c$ Service de Physique de l'Univers, Champs et Gravitation, Université de Mons -- UMONS, Place du Parc 20, 7000 Mons, Belgium}

\emailAdd{mkarydas@berkeley.edu} \emailAdd{songyuanli123@gmail.com}
\emailAdd{petkou@auth.gr}
\emailAdd{matthieu.vilatte@umons.ac.be}

\abstract{We present the details of a recently discovered representation of conformal four-point ladder integrals as thermal one-point functions in scalar field theories. We show that the conformal ladder integrals can be constructed from the partition function of two harmonic oscillators twisted by an imaginary chemical potential and that for any even dimension $D$ and any loop order $L$ they satisfy a familiar second order differential equation. In our representation, thermal one-point functions of higher-spin operators correspond to linear combinations of multi-loop ladder graphs in $D=2$ and $D=4$ dimensions.  Moreover, we give a simple derivation for the all-loop resummation of conformal ladder integrals for arbitrary $D$.  We conclude by highlighting possible connections between our work and recent developments in the thermal bootstrap, multiloop calculations, integrability, AdS/CFT and string theory.}

\begin{document}
\maketitle
\flushbottom

\section{Introduction}
\label{sec:intro}

Four-point functions are quantities of prime interest in  conformal field theories (CFTs) in any dimension as, under a few general assumptions, their knowledge leads to the complete solution of the theory. Although one might aim to evaluate them using only the symmetries of the CFT, it is useful to keep in mind that they can also be calculated perturbatively using traditional Feynman integral methods. The first truly conformal technique for the calculation of $n$-point Feynman integrals with $n\geq 3$ was presented in Symanzik's important work \cite{Symanzik:1972wj}, however, it took nearly twenty years before it was applied to explicit calculations in realistic CFT models \cite{Petkou:1994ad,Hoffmann:2000mx,Hoffmann:2000tr}. With the advent of AdS/CFT, Symanzik's technique was rediscovered and refined several times (see e.g. \cite{Dolan:2000ut}) and it has since been widely applied in CFT calculations.\footnote{For some recent development see \cite{Alkalaev:2025zhg,Alkalaev:2025fgn}.}

In an apparently unrelated development, Mellin transform techniques were used to calculate multiloop four-point ``ladder" integrals \cite{Usyukina:1992jd,Usyukina:1993ch}. The relevance of conformal symmetry to those important calculations was emphasised in \cite{Broadhurst:1993ib,Broadhurst:2010ds}. Again, AdS/CFT reignited the interest in such results and led to a remarkable progress in calculational techniques that extend much beyond CFTs and four-point functions.\footnote{It is hard to include here the immense literature on multiloop Feynman integrals. A recent review with extended bibliography is \cite{Abreu:2022mfk}. } 
Two highlights of the recent progress in multiloop calculations are the connection to the theory of single-valued polylogarithms (see e.g. \cite{Brown:2004ugm,Schnetz:2013hqa}) and to integrability (see e.g. \cite{Isaev:2003tk,Derkachov:2021ufp}). 

Nevertheless, conformal four-point integrals continue to surprise us with their properties and their connections to various quantities in physics and mathematics. In this work we will discuss in detail a novel relationship between a class of conformal four-point integrals and some at first sight unrelated quantities - the thermal one-point functions of higher-spin composite operators in massive free QFTs. We call this the {\it thermal representation of conformal ladder integrals}. Our observation has its origins in \cite{Petkou:2018ynm}, which in turn was motivated by efforts to generalise the conformal bootstrap technique to finite temperature \cite{Iliesiu:2018fao}. In the present work we build upon preliminary results presented in \cite{Petkou:2021zhg,Karydas:2023ufs}, with the aim of synthesizing and extending them.

The paper is organized as follows: in Section \ref{sec: conf ladder integrals} we review the properties and the integral representations of conformal ladder integrals in $D=2k+2$ dimensions for any loop order $L$. In Section \ref{sec: QM model} we present our parent quantum mechanical model of two twisted harmonic oscillators  which we use to construct the thermal free energies of ideal relativistic gases in $d=2L+1$ dimensions. We show that these free energies are given by iterated integrals and satisfy various differential recursive relations. We also discuss a ``thermal" representation for massless two-point functions of scalar fields in any dimension $D=2k+2$. In Section \ref{sec: correspondence} we present our {\it thermal representation of conformal ladder integrals} and show how the latter are built from the free energy of our parent quantum mechanical model via the successive application of a set of integrodifferential operators. We also prove that the conformal ladder integrals satisfy a familiar second order differential equation. In Section \ref{sec: applications} we discuss two applications of our formalism. The first concerns the representation of thermal one-point functions of higher-spin conserved currents as linear combinations of multi-loop conformal ladder integrals in $D=2$ and $D=4$ dimensions. The second is the introduction of a {\it hyper-partition function} which encodes the all-loop resummation of conformal ladder integrals for any $D$. In Section \ref{sec: conclusion} we summarize our results and indicate a number of possible connections of our work to a range of active research directions. Technical issues are left in the five Appendices.

\section{Conformal ladder integrals in $D=2k+2$  dimensions}
\label{sec: conf ladder integrals}

We consider the following class of four-point $L$-loop $(L\geq 0)$ conformal ladder integrals in even dimensions $D=2k+2$, $k= 1,2,...$ 
\be
\label{ILk_def}
g^{2L} I_L^k(x_1,x_2,x_3,x_4)=\prod_{n=1}^{L}\left[\int\frac{d^{2k+2} y_n}{\pi^{{k+1}}}\frac{g^2\Gamma(k)}{|y_{n-1,n}|^{2k}|x_2-y_n|^2|x_4-y_n|^2}\right]\frac{1}{|y_L-x_3|^{2k}}\,,
\ee
with $y_0=x_1$ and $y_{ij}=y_i-y_j$.\footnote{ For $L=0$ we have $I_0^k= 1/|x_1 - x_3|^{2k}$.} We have also introduced the dimensionless loop-counting parameter $g^2$ and the factor $\Gamma(k)$ in the numerators of the r.h.s. for later convenience with the formulae. For $k=1$ ($D=4$) such integrals are omnipresent in multiloop calculations of ${\cal N}=4$ SYM \cite{Gromov:2018hut} and in studies of integrability and Yangian symmetry \cite{Loebbert:2022nfu}. For general $2\leq D\leq 4$ such integrals appear in the generalised fishnet models introduced in \cite{Kazakov:2018qbr} and more recently  in large charge calculations  \cite{Giombi:2020enj,Giombi:2022gjj,Caetano:2023zwe}. The singular case $D=2$ ($k=0$) was considered in \cite{Derkachov:2018rot} and it is reviewed in Appendix \ref{App:A}. Using standard normalizations of massless propagators\footnote{To fix conventions we note that the $x$-space two-point function of the scalar field $\phi$ with Lagrangian ${\protect
  \cal L}=(1/2)\phi (-\partial ^2)^a\phi $ in $d=2L+1$-dimensions is $C_\phi
  ^L(a)/x^{2L+1-2a}$, with $C_{\phi }^L(a)=\Gamma (L+1/2-a)/\Gamma (a)4^a\pi
  ^{L+1/2}$.} one can obtain the relation between the loop-counting parameter $g^2$ and the dimensionless coupling of the corresponding CFT model that gives rise to the integrals, however we will not do this here as our discussion is more general. 

\begin{figure}[h!]
    \centering    \includegraphics[width=0.9\linewidth]{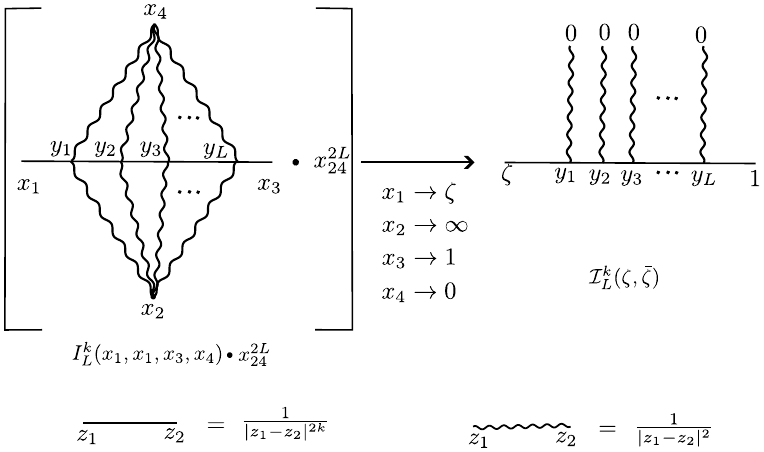}
    \caption{Graphical representation of $\mathcal{I}^{k}_{L}(\z, \bar{\z})$}
    \label{fig: conf integrals}
\end{figure}

Apart from the obvious Lorentz and translation invariance these integrals transform under dilatations and inversions of the coordinates $x_i,i=1,2,3,4$ as 
\begin{align}
\label{ILk_dil}
I_L^k(\lambda x_1,\lambda x_2,\lambda x_3,\lambda x_4)&=\frac{1}{\lambda^{2k+2L}}I_L^k(x_1,x_2,x_3,x_4)\,,\\
\label{ILk_inv}
I_L^k\left(\frac{1}{x_1},\frac{1}{x_2},\frac{1}{x_3},\frac{1}{x_4}\right)&=(x_{1}^2x_3^2)^k(x_2^2x_4^2)^L I_L^k(x_1,x_2,x_3,x_4)\,,
\end{align}
with $(1/x_i)_\m=(x_{i})_{\m}/x_i^2$. Namely, they correspond to conformally covariant four-point functions of weights $k$ and $L$ in the channels $(1-3)$ and $(2-4)$ respectively. Consequently they can be represented by a single function $\Phi(u,v)$ of the conformal ratios $u,v$ as
\be
\label{ILk_gen}
I_L^k(x_1,x_2,x_3,x_4)=\frac{1}{x_{13}^{2k}x_{24}^{2L}}\Phi_L^k(u,v)\,,\,\,\,u=\frac{x_{12}^2x_{34}^2}{x_{13}^2x_{24}^2}\,,\,\,\,v=\frac{x_{12}^2x_{34}^2}{x_{14}^2x_{23}^2}\,,
\ee
where again $x_{ij} = x_{i} - x_{j}$. As usual, the representations of conformal four-point functions can be further simplified if we use conformal symmetry to fix the positions of the $\{x_i\}$ as $x_1\mapsto \ze$, $x_2\mapsto \infty$, $x_3\mapsto 1$ and $x_4\mapsto 0$ in which case we can write (\ref{ILk_gen}) as 
\be
\label{ILk_lim}
\lim_{\{x_i\}\mapsto \{\ze,\infty,1,0\}}[x_{24}^{2L} \,I_L^k(x_1,x_2,x_3,x_4)]\equiv \cI_L^k(\ze,\bze)=\frac{1}{|1-\ze|^{2k}}\Phi_L^k(\ze,\bze)\,,
\ee
where $\ze$ and $\bze$ are defined by 
\be
\label{uvzbz}
\left(u,v\right)=\left(\frac{1}{(1-\ze)(1-\bze)},\frac{1}{\ze\bze}\right)\,.
\ee
The function $\Phi_L^k(\ze,\bze)$ is the main object of interest in four-point calculations as it encodes all the relevant dynamical information of the corresponding CFT. 
A frequently used graphical representation for the functions $\cI_L^k(\ze,\bze)$ is shown in Fig. \ref{fig: conf integrals}.

Since their original evaluation in \cite{Usyukina:1992jd,Usyukina:1993ch} (see also \cite{Broadhurst:1993ib, Broadhurst:2010ds}) the functions $\cI_L^1(\ze,\bze)$, corresponding to conformal $L$-loop ladder integrals in $D=4$, have appeared in numerous calculations of multiloop Feynman integrals. Notably, they are the building blocks for the Basso-Dixon integrals \cite{Basso:2017jwq,Basso:2021omx} giving the large-$N_c$ limit of the fishnet theories  \cite{Gurdogan:2015csr,Kazakov:2018qbr,Kazakov:2022dbd}. The functions $\cI_L^0(\ze,\bze)$ correspond to conformal integrals in $D=2$ dimensions. As we already stressed this is a singular limit that requires care and it was discussed in the context of a two-dimensional fishnet CFT in \cite{Derkachov:2018rot} and it is reviewed in Appendix \ref{App:A}.  Recently it was suggested that the two dimensional conformal integrals are related to Calabi-Yau geometry \cite{Duhr:2022pch,Duhr:2023eld,Duhr:2023bku,Duhr:2024hjf}. 

In a remarkable result  Isaev has given integral representations for the functions $\cI_L^k(\ze,\bze)$ for $k=1,2,3,..$ using an algebraic approach based on conformal quantum mechanics \cite{Isaev:2003tk}. These results have been recently used in large charge calculations \cite{Giombi:2020enj,Giombi:2022gjj,Caetano:2023zwe,Brown:2025cbz}. Isaev's representation for the functions $\cI_L^k(\ze,\bze)$ reads (i.e. eq. (47) of \cite{Isaev:2003tk} with the appropriate normalisations)
\begin{align}
\label{ILk_Isaev}
\cI_L^k(\ze,\bze)&=-\frac{1}{(L!)^2} \int_0^\infty dt \, t^L[t-2\ln|\z|]^L \partial_t\left(\frac{e^{-t}}{(1-\ze e^{-t})(1-\bze e^{-t})}\right)^k\,,\nonumber\\
&=\frac{1}{L!(L-1)!} \int_0^\infty dt \, 2(t-\ln|\ze|)[t([t-2\ln|\ze|])]^{L-1} \frac{e^{-kt}}{|1-\ze e^{-t}|^{2k}}\,.
\end{align}
The second line has been obtained by partial integration assuming $L\geq 1$. 
 Using the change of variables 
\be
\label{zz'}
\ze'=\ze e^{-t}\,,\,\,\,\ze=|\ze|e^{i\theta}\,,\,\,\ze'=|\ze'|e^{i\theta}\,,
\ee
we obtain from (\ref{ILk_lim}) and (\ref{ILk_Isaev})
\be
\label{ILk_our1}
\Phi_L^k(\ze,\bze)=\frac{1}{\cD_0^k(\ze,\bze)}\frac{1}{L!(L-1)!}\int_0^{|\ze|}\frac{d|\ze'|}{|\ze'|}2\ln|\ze'|(\ln^2|\ze'|-\ln^2|\ze|)^{L-1}\cD_0^k(\ze',\bze')\,,
\ee
where we have defined for later use the functions
\be
\label{D0k}
\cD_0^k(\ze,\bze)=\Gamma(k)\frac{(\ze-\bze)^k}{|1-\ze|^{2k}}\,,\,\,\, k=1,2,3,..\,.
\ee
The case $k=0$ ($D=2$) can be included in the above formula if we define
\be
\label{D00}
\cD_0^0(\ze,\bze)\equiv\ln\cZ_0(\ze,\bze)\,,
\ee
where $\cZ_0(\ze,\bze)$ is the partition function (\ref{defF}) of the quantum mechanics model that we discuss in the next section. Notice that with our definitions $\Phi_0^k(\z,\bze)=1$. 
 
The representation (\ref{ILk_our1}) of the conformal ladder graphs for all loops and (even) dimensions is a straightforward extension of the ones given in \cite{Isaev:2003tk}. Yet, as we will show below, written in this way it unveils a remarkable connection with the thermal free energies of massive scalar field theories.

\section{Thermal free energies of ideal relativistic gases and massive complex scalars}
\label{sec: QM model}

One of the main result of this work is to show that the conformal four-point ladder integrals (\ref{ILk_def}) and (\ref{ILk_our1}) are built from the free energy of the simple quantum mechanical model that involves two harmonic oscillators - the  {\it parent system}.

\subsection{The parent quantum mechanical system}

The parent system consists of pair of {\it twisted} harmonic oscillators. The twisting is a particular coupling between the oscillators that involves an imaginary parameter. This can be viewed as an interaction with an imaginary magnetic field (see e.g. \cite{Daoud:2009uj}), or the introduction of an imaginary chemical potential. However we will use here the term ``twisting" to emphasize the effect of such a coupling to the Hilbert space of the system in the spirit of \cite{Dunne:2018hog}. The parent system Hamiltonian is  \be
\label{BasicHamiltonian}
\hat{H}=\frac{1}{2}\hat{p}^2_1+\frac{1}{2}\hat{p}^2_2 +\frac{1}{2}m^2(\hat{x}^2_1+ \hat{x}^2_2)+i\mu\left(\hat{p}_{2}\hat{x}_1-\hat{p}_{1}\hat{x}_2\right)\,,
\ee
where $m$ is the common frequency/mass, $\m$ is the twisting parameter and we set $\hbar=1$ onwards. 
The commutation relations are as usual $[\hat{x}_i,\hat{p}_j]=i\delta_{ij}$. This Hamiltonian can be thought of as a deformation of  the free particle Hamiltonian
\be
\label{H0}
\hat{H}_0=(\hat{p}^2_1+\hat{p}^2_2)/2\,,
\ee
by the operators 
\be
\label{O0Q0}
\hat{\cO}=\frac{1}{2}(\hat{x}^2_1+ \hat{x}^2_2)\,,\,\,\,\,\hat{\cQ}=\hat{p}_{2}\hat{x}_1-\hat{p}_{1}\hat{x}_2\,,
\ee
the deformations parameters being $m$ and $i\mu$. We can disentangle the system defining the creation/annihilation operators as
\begin{align}
\label{a1}
\hat{a}^\dagger_{1}&=\frac{1}{2\sqrt{m}}(m(\hat{x}_{1}+i\hat{x}_{2})+ (\hat{p}_{2}-i\hat{p}_{1}))\,,\,\,\,\hat{a}_{1}=\frac{1}{2\sqrt{m}}(m(\hat{x}_{1}-i\hat{x}_{2})+ (\hat{p}_{2}+i\hat{p}_{1}))\,,\\
\label{a2}
\hat{a}^\dagger_{2}&=\frac{1}{2\sqrt{m}}(m(\hat{x}_{1}-i\hat{x}_{2})- (\hat{p}_{2}+i \hat{p}_{1}))\,,\,\,\,\hat{a}_{2}=\frac{1}{2\sqrt{m}}(m(\hat{x}_{1}+i\hat{x}_{2})- (\hat{p}_{2}-i \hat{p}_{1}))\,,
\end{align}
which satisfy
\be
\label{commrel}
[\hat{a}_i,\hat{a}_j^\dagger]=\delta_{ij}\,,\,i=1,2\,.
\ee
It is then straightforward to obtain
\begin{align}
\label{Ha1}
\hat{H}_0+m^2\hat{\cO} &=m(\hat{\bf N}_1+\hat{\bf N}_2+1) \,,\\
\hat{\cQ} &=\hat{\bf N}_1-\hat{\bf N}_2\,,
\end{align}
with the number operators  defined as ${\hat{\bf N}}_i=\hat{a}^\dagger_i \hat{a}_i$ for $i=1,2$. Therefore the Hilbert space\footnote{The careful reader will notice that the Hilbert space of our system can be mapped to the $SU(2)$ irreps via the Schwinger boson construction. 
} of the system is the tensor product space ${\cal H}_{1,2}\approx \{|n_1\rangle\otimes|n_2\rangle\}$ with $n_1,n_2=0,1,2,..$ the eigenvalues of the corresponding number operators $\hat{\bf N}_i$. 
The above facilitate the calculation of  the grand canonical partition function of the system  as
\begin{align}
\label{Z0def}
{\cal Z}_0(z,\bar{z})&=\Tr_{{\cal H}_{1,2}} \left[e^{-\beta (\hat{H}_0+m^2\hat{\cO})}e^{-i\beta\mu\hat{\cQ}}\right]=\sum^\infty_{n_1,n_2=0}e^{-m\beta(n_1+n_2+1)-i\mu\beta(n_1-n_2)}\,,\nonumber\\
&=e^{-\beta m}\sum_{n=0}^\infty  z^n\sum_{m=0}^\infty \bz^m=e^{\ln|z|-\ln(1-z)-\ln(1-\bz)}\,,
\end{align}
where the variable $z$ is defined as
\be
\label{z}
z=e^{-\beta m-i\beta\mu}\,.
\ee
The building block in our construction of conformal ladder graphs is the free energy $\cF_0(z,\bz)$ of the parent quantum mechanical system defined as
\be
\label{defF}
\ln \cZ_0(z,\bz)=-\beta \cF_0(z,\bz)=\ln |z|-\ln(1-z)-\ln(1-\bz)\,.
\ee
Of central importance for what follows are the two differential operators
\begin{align}
\label{opD}
\hat{\bf D}_z&=\frac{1}{\beta^2}\frac{\partial}{\partial m^2}=\frac{1}{2\ln|z|}(z\partial_z+\bz\partial_\bz)\,,\\
\label{opL}
\hat{\,\bf L}_z&=\frac{i}{\beta}\frac{\partial}{\partial\mu}=(z\partial_z-\bz\partial_\bz)\,,
\end{align}
which commute $[\hat{\bf D}_z,\hat{\,\bf L}_z]=0$. A useful result is that the Laplacian in the variables $m$ and $\mu$ can be written as
\begin{equation}
\label{Laplacian}
 \hat{\bf \Delta}_z=\frac{\partial^2}{\partial m^2}+\frac{\partial^2}{\partial\mu^2}=   4\b^2\,z\bz\partial_z\partial_{\bz}=4\b^2\left(\ln^2|z|\hat{\bf D}_z^2+\frac{1}{2}\hat{\bf D}_z-\frac{1}{4}\hat{\,\bf L}_z^2\right)\,.
 \end{equation} 
Using (\ref{opD}) and (\ref{opL}) we obtain the thermal one-point functions of  $\hat{\cal O}$ and $\hat{\cal Q}$ as\footnote{We use the symbol $*$ to denote the action of a differential or integral operator on a function.}
\begin{align}
\label{eq: Oaverage}
\langle\hat{\cal O}(z,\bz)\rangle_0&=-\beta\hat{\bf D}_z*\ln \cZ_0(z,\bz)=\frac{1}{2m}\langle\hat{\bf N}_1+\hat{\bf N}_2+1\rangle_0=\frac{1}{2m}\frac{1-|z|^2}{|1-z|^2}\,,\\
\label{eq:Qaverage}
\langle\hat{\cal Q}(z,\bz)\rangle_0 &=\hat{\,\bf L}_z*\ln \cZ_0(z,\bz)=\langle\hat{\bf N}_1-\hat{\bf N}_2\rangle_0=\frac{z-\bz}{|1-z|^2}\,.\end{align}
Interestingly, these are related to Poisson kernels: $\langle\hat{\cal O}(z,\bz)\rangle_0$ and $\langle\hat{\cal Q}(z,\bz)\rangle_0$ are harmonic functions on the open unit disk  and on the upper half plane respectively, that take constant values on their corresponding boundaries. 

\subsection{From the parent system to ideal relativistic gases ...}
Viewing  $\cZ_0(z,\bz)$ as the grand canonical partition function of a pair of ``photons" with common energy $m$ subject to opposite ``imaginary" chemical potential $\mu$, we can calculate the free energy of an ideal relativistic gas in $d=2L+1$ dimensions by integrating $\ln\cZ_0(z,\bz)$ with a suitable one-particle density of states.\footnote{Integrals such as the ones considered here, for real chemical potential, have been considered in the past among others in \cite{Haber:1981fg,Haber:1981tr,Haber:1981ts}. The case of $d$ even is more complicated and we have not found any useful diagrammatic correspondence for it. Nevertheless, it deserves to be studied further.} 
The latter  
is found by considering a system of non-interacting bosons with relativistic energies $\omega^2=\vec{p}^2+m^2$ placed in a $(d-1)$-dimensional spatial cubic box of volume $V_{d-1}=\ell^{d-1}$ with quantized  spatial momentum  $\vec{p}=\left(\frac{2\pi}{\ell}n_1,...,\frac{2\pi}{\ell}n_{d-1}\right)=\frac{2\pi}{\ell}\vec{n}
$. The number of modes having momenta inside the spherical shell bounded by $|\vec{p}|$ and $|\vec{p}|+d|\vec{p}|$ is 
\be\label{dn}
dn=\left(\frac{\ell^2}{4\pi^2}\right)^L |\vec{p}|^{2L-1}d|\vec{p}| d\Omega_{2L}\,,
\ee
with $\int d\Omega_{2L}=2\pi^L/\Gamma(L)$.
Using the above we find the one-particle density of states as
\begin{equation}
    \label{rho_L}
    \rho_L(\omega;m;\a^2) = \frac{dn}{d\omega}=\frac{2\a^{2L}\b^{2L}}{(L-1)!}\omega (\omega^2-m^2)^{L-1}\,,
\end{equation}
where we have defined the dimensionless {\it geometric parameter}
\be
\label{alpha}
\alpha^2=\frac{\ell^2}{4\pi\beta^2}\,.
\ee
To obtain the sought-after free energy $\ln \cZ_L(z,\bz;\a^2)$ for $L\geq 1$ we integrate over the one-particle density of states as
\begin{align}
\label{Zdexpl}
\ln\cZ_L(z,\bz;\a^2)&=\frac{\a^{2L}\b^{2L}}{(L-1)!}\int_m^\infty d\omega \,2\omega (\omega^2-m^2)^{L-1}\ln \cZ_0(z',\bz')\nonumber \\
&=-\frac{\a^{2L}}{(L-1)!}\int_0^{|z|}\frac{d|z'|}{|z'|}2\ln|z'|(\ln^2|z'|-\ln^2|z|)^{L-1}\ln\cZ_0(z',\bz')\,,
\end{align}
where $z'=e^{-\b\omega-i\b\m}$ such that $\ln|z'|<\ln|z|$. The divergence for  $\omega\rightarrow\infty$ (or $|z'|\rightarrow  0$) corresponds to the contribution of the vacuum energies and can be subtracted to obtain a finite result. 

We can already notice the relationship between the thermal free energies above and the conformal ladder graphs comparing the second line of  (\ref{Zdexpl}) with the  representation (\ref{ILk_our1}). We also see that $\a^2$ plays the role of a {\it dimension-counting} parameter for the thermal free energies, and we can define 
\be
\label{ZLalpha}
\ln\cZ_L(z,\bz;\a^2)=\a^{2L}\ln\cZ_L(z,\bz)\,.
\ee
The explicit evaluation of (\ref{Zdexpl}) was done in \cite{Petkou:2021zhg} and gives (see also Appendix \ref{App:B})
\begin{mybox}
\be
\label{ZLresult}
    \ln \cZ_L(z,\bz)=\frac{(-1)^LL!(2\ln|z|)^{2L+1}}{2(2L+1)!}+\sum_{n=0}^{L} \frac{ (2L-n)! (-2\ln |z|)^n}{(L-n)!n!} 2\Re[Li_{2L+1-n}(z)]\,,
\ee
\end{mybox}
 where $Li_s(z)$ are the standard polylogarithms see e.g. \cite{Goncharov:1995tdt}. Notice that for $m=\m=0$ we have
 \be
 \label{ZL0}
 \ln\cZ_L(1,1)=\frac{(2L)!}{L!}2\zeta(2L+1)\,,
 \ee
 which diverges for $L=0$ and it is finite otherwise. In Appendix \ref{App:E} we provide an alternative expression for $\ln\cZ_L(z,\bz)$ in term of the Zagier-Ramakrishnan single-valued polylogarithms \cite{Zagier2,Goncharov:1996dce}, which are sometimes useful in the calculation of holographic correlators e.g. \cite{Ceplak:2021wzz}.

\subsection{... and to free massive complex scalars}

As shown  in \cite{Petkou:2021zhg} and briefly reviewed in Appendix \ref{App:B},  (\ref{ZLresult}) coincides with the standard calculation of the logarithm of the thermal partition function  for a free massive complex scalar field $\phi(x)$ in the presence of imaginary chemical potential, or equivalently an appropriately chosen real background gauge potential for a $U(1)$ charge  \cite{Filothodoros:2016txa,Filothodoros:2018pdj}. The Euclidean action is
\be
\label{SE}
{\cal S}_L(\b;m,\m)=\int_0^\beta d\tau\int d^{2L}\vec{x} \,\,|(\partial_\tau-i\m)\phi|^2+|\vec{\partial}\phi|^2+m^2|\phi|^2\,,
\ee
and the regularized\footnote{The regularization consists of dividing with the zero temperature quantity $\cZ(\infty;0,0)$ which subtracts the zero-point energies of the harmonic oscillators when we calculate $\mathcal{F}_{L}(\b;m,\m)$,} grand canonical partition function reads 
 \be
\label{Zgcphi}
\cZ_{L}(\b;m,\m)\equiv\frac{1}{\cZ(\infty;0,0)}\int ({\cal D}\bphi)({\cal D}\phi)e^{-S_E}=e^{-\b \mathcal{F}_{L}(\b;m,\m)}\,.
\ee
The path integral is Gaussian and can be calculated using standard methods. The result is proportional to (\ref{ZLresult}), and to match it with the ideal relativistic gas calculations of $\ln \cZ_L(z,\bz)$ we need to identify the dimensionless parameter $\a^{2L}$ in (\ref{alpha}) with the ratio $V_{2L}/\b^{2L}$, where $V_{2L}$ is the spatial volume of ${\mathbb R}^{2L}$ regularised by some length scale proportional to $\ell$.

Having at hand the partition functions $\cZ_L(z,\bz)$ we can use the operators $\hat{\bf D}_z$ and $\hat{\,\bf L}_z$ defined in (\ref{opD}) and (\ref{opL}) to calculate the thermal averages of the operators $\cal O$ and $\cal Q$ 
\be
\label{eq:OQ}
\cO=|\phi|^2\,,\,\,\,\,\,\cQ=\phi^\dagger\overleftrightarrow{{\cal D}_\tau}\phi\,,\,\,\,\,{\cal D}_\tau=\partial_\tau-i\mu\,.
\ee
In doing so we find that the action of the first order differential operator $\hat{\bf D}_z$ on $\ln\cZ_L(z,\bz)$ not only gives the one-point function $\langle \cO(z,\bz)\rangle_L$, but at the same time it yields $\ln\cZ_{L-1}(z,\bz)$. In other words,  the thermal one-point function $\langle\cO(z,\bz)\rangle_L$ in $d=2L+1$ dimensions corresponds  the  thermal free energy of the massive free theory in $d-2$ dimensions. We summarize below the results\footnote{Differential equation satisfied by conformal integrals were studied in $x$-space some time ago e.g.  \cite{Drummond:2010cz,Drummond:2012bg} and it would be nice to understand their relation to the ones derived here.}  of \cite{Petkou:2021zhg}
\begin{align}
\label{Oaverage}
&\langle\cO(z,\bz)\rangle_L=-\beta\,\hat{\bf D}_z*\ln \cZ_L(z,\bz)=\b\ln \cZ_{L-1}(z,\bz)\\
\label{Qaverage}
&\langle\cQ(z,\bz)\rangle_L=\hat{\,\bf L}_z*\ln \cZ_{L}(z,\bz)=-\hat{\bf D}_z*\langle\cQ(z,\bz)\rangle_{L+1}\,,
\end{align}
for $L=0,1,2,3,..$. Notice that we have defined $ \beta\ln \cZ_{-1}(z,\bz)\equiv\langle{\cal O}(z,\bz)\rangle_0$ whose explicit expression was given in \eqref{eq: Oaverage}.

\subsection{The iterated integral representation}

The thermal free energies above can be represented as iterated integrals. Using the result
\be
\label{A1}
(\ln^2|z_1|-\ln^2|z|)^{L-1}=-(L-1)\int_{|z_1|}^{|z|}\frac{d|z_2|}{|z_2|}2\ln|z_2|(\ln^2|z_1|-\ln^2|z_2|)^{L-2}\,,
\ee
we can write (\ref{Zdexpl}) as
\be
\label{A2}
\ln\cZ_L(z,\bz)=\frac{1}{(L-2)!}\int_0^{|z|}\frac{d|z_1|}{|z_1|}2\ln|z_1|\int_{|z_1|}^{|z|}\frac{d|z_2|}{|z_2|}2\ln|z_2|(\ln^2|z_1|-\ln^2|z_2|)^{L-2}\ln\cZ_0(z_1,\bz_1)\,.
\ee
Notice that the integration variables satisfy $0\leq|z_1|\leq|z_2|\leq |z|$ and this allows us to reverse the order of integration as
\be
\label{A3}
\ln\cZ_L(z,\bz)=\frac{1}{(L-2)!}\int_0^{|z|}\frac{d|z_2|}{|z_2|}2\ln|z_2|\int_{0}^{|z_2|}\frac{d|z_1|}{|z_1|}2\ln|z_1|(\ln^2|z_1|-\ln^2|z_2|)^{L-2}\ln\cZ_0(z_1,\bz_1)\,.
\ee
One can verify this by acting on (\ref{A2}) and (\ref{A3}) with the differential operator $\hat{\bf D}_z$ to obtain the same result. We can use again (\ref{A1})  inside (\ref{A3}) introducing a new integration variable $z_3$ where now $0\leq|z_1|\leq |z_3|\leq|z_2|\leq |z|$. Hence, we can move the integral over $z_3$ to the left of the integral over $z_1$. We can continue this procedure $L-1$ times to obtain 
\begin{mybox}
\be
\label{ZL}
    \ln \cZ_L(z,\bz)=(-1)^L\left\{{\bf \text{ord}}\prod_{i=1}^{L}\right\}\left[\int_0^{|z_{i+1}|}\frac{d|z_i|}{|z_i|}2\ln |z_i|\right]\ln \cZ_0(z_1,\bz_1)\,,
\ee
\end{mybox}
where $z_i=|z_i|e^{-i\beta\mu}$, $z_{L+1}=z$. We used the {\it ordered product} symbol $\left\{\text{ord}\prod_{i=1}^L\right\}$ to denote an iterated integral that is performed with the variables ordered from right to left as $0\leq |z_1|\leq |z_2|\leq ..\leq |z_L|\leq |z|$. Writing (\ref{defF}) as
\begin{equation}
\label{Z0}
    \ln \cZ_0(z,\bz)=\int_0^{z}\frac{dz'}{1-z'}+\int_0^{\bz}\frac{dz'}{1-z'}-\int_{|z|}^1\frac{dz'}{z'}\,,
\end{equation}
we see that (\ref{ZL}) coincides with the class of iterated integrals that give rise to single-valued polylogarithms \cite{Brown:2004ugm, Schnetz:2013hqa}.

The  representation (\ref{ZL}) of $\ln \cZ_L(z,\bz)$ motivates the introduction of the integral operator $ \check{\!\mathbf{d}}_{z;z'}$  defined as\footnote{More formally, we may consider the differential operators $\hat{\bf D}_z$ and $\hat{\,\bf L}_z$ as a set of orthogonal vectors on the two dimensional Euclidean space with complex coordinates $z,\bar{z}$ and metric $ds^2=dzd\bar{z}$. The corresponding dual one-forms are 
\be
\label{1forms}
\check{\bf D}_z=\ln|z|\left(\frac{dz}{z}+\frac{d\bar{z}}{\bar{z}}\right)\,,\,\,\,\,\check{\,\bf L}_z=\frac{1}{2}\left(\frac{dz}{z}-\frac{d\bar{z}}{\bar{z}}\right)\,,
\ee
such that the inner products are
\be
\label{InnerProducts}
\langle \check{\bf D}_z,\hat{\bf D}_z\rangle=1\,,\,\,\langle \check{\bf D}_z,\hat{\,\bf L}_z\rangle=0\,,\,\,\,\langle \check{\,\bf L}_z,\hat{\bf D}_z\rangle=0\,,\,\,\langle \check{\,\bf L}_z,\hat{\,\bf L}_z\rangle=1\,.
\ee}

\be
\label{d_def}
\check{\!\mathbf{d}}_{z;z'}=\int_0^{|z|}\frac{d|z'|}{|z'|}2\ln|z'|\,,
\ee
with the following properties
\begin{align}
\label{dD}
\check{\!\mathbf{d}}_{z;z'}*\hat{\bf D}_{z'}*f(z')&=\hat{\bf D}_z*\check{\!\mathbf{d}}_{z;z'}*f(z')=f(z)\,,\\
\label{Ld}
\check{\!\mathbf{d}}_{z;z'}*\hat{\,\bf L}_{z'}*f(z')&= \hat{\,\bf L}_z*\check{\!\mathbf{d}}_{z;z'}*f(z')\,,
\end{align}
assuming that $f(0)=0$. Then (\ref{ZL}) can be written more compactly as\begin{align}
\label{ZLd}
\ln \cZ_L(z,\bz)&=(-1)^{L}\,\check{\!\mathbf{d}}_{z;z_{L}}*...*\check{\!\mathbf{d}}_{z_3;z_2}*\check{\!\mathbf{d}}_{z_2;z_1}*\ln \cZ_0(z_1,\bz_1) \equiv [-\,\check{\!\mathbf{d}}]^L_{z;z'}*\ln \cZ_0(z',\bz')\,,
\end{align}
which serves as a definition for the exponentiation of the operator $[\,\check{\!\mathbf{d}}]_{z;z'}^{L}$.

\subsection{A ``thermal" representation for massless two-point functions in $D=2k+2$ dimensions}

The above formalism yields a ``thermal" representation for the massless two-point functions of elementary scalar fields in even dimensions $D=2k+2$, $k=0,1,2,...$. Indeed, we can verify by a straightforward calculation that
\be
\label{ellOp}
(z-\bz)^k\left(\frac{1}{z-\bar{z}}\hat{\,\bf L}_z\right)^k*\ln \cZ_0(z,\bz)=\cD_0^k(z,\bz)\,,
\ee
with $\cD_0^k(z,\bz)$ defined in (\ref{D0k}). Setting $D=2k+2$ we see that the r.h.s. of (\ref{ellOp}) is related to the massless propagator of a scalar field $\phi(x)$ with scaling dimension $\Delta_\phi=D/2-1 = k$ in $D$-dimensions. For $k=0$ we note that $\ln \cZ_0(z,\bz)$ is related to the two-point functions of the (non-primary) massless scalar in $D=2$. The representation (\ref{ellOp}) implies that  massless two-point functions in $D$-dimensions can be viewed as a thermal cumulants in our parent quantum mechanical system (see Appendix \ref{app:cumulants} for more details). We call the positive integer $k$ the {\it depth}.  

The functions $\cD_0^k(z,\bz)$ satisfy the recursive relation 
\be
\label{Lzk}
\left(\hat{\,\bf L}_z-k\frac{z+\bz}{z-\bz}\right)*\cD_0^k(z,\bz)=\cD_0^{k+1}(z,\bz)\,,\,\,\,k=0,1,2,3,... \, .
\ee
In other words the operator 
\be
\label{Lzk_def}
\hat{\,\bf L}_z^{(k)}\equiv\hat{\,\bf L}_z-k\frac{z+\bz}{z-\bz}\,,
\ee
raises the value of the depth parameter $k\mapsto k+1$ or equivalently raises the dimension $D\mapsto D+2$. It also commutes with $\check{\bf d}_{z;z'}$ as
\be
\label{Lzkd}
\check{\!\mathbf{d}}_{z;z'}*\hat{\,\bf L}^{(k)}_{z'}*f(z')= \hat{\,\bf L}^{(k)}_z*\check{\!\mathbf{d}}_{z;z'}*f(z')\,.
\ee
The above motivate us to write\footnote{Note the difference between the operator $\hat{\,\bf L}_z^{(k)}$ and the exponentiation $[\hat{\,\bf{L}}_z]^k$.}
\be
\label{LzK_def}
\left\{\text{ord}\prod_{n=0}^{k-1}\right\}\left[\hat{\,\bf L}_z-n\frac{z+\bz}{z-\bz}\right]*\ln\cZ_0(z,\bz)\equiv [\hat{\,\bf{L}}_z]^k*\ln\cZ_0(z,\bz)=\cD_0^k(z,\bz)\,.
\ee
As in \eqref{ZL} we have used again the ordered product symbol to indicate that $n$ increases from right to left. We further notice that $\hat{\,\bf L}_z^{(k)}$ defined in (\ref{Lzk_def}) commutes with $\hat{\bf D}_z$
\be
\label{LzkD}
[\hat{\,\bf L}_z^{(k)},\hat{\bf D}_z]=0\,.
\ee
We summarize in Tables \ref{tab:my_label_1}, \ref{tab:my_label_2}  and \ref{tab:my_label_3}  the definitions of the various operators defined above, their properties and their action on the conformal integrals free energies $\ln \mathcal{Z}_{L} (z,\bz)$ or $\cD_0^k(z,\bz)$.

\begin{table}[h!]
    \centering
    \renewcommand{\arraystretch}{1.3}
    \begin{tabular}{>{$}l<{$} >{$}l<{$} >{$}l<{$}}
    \toprule
    \ln\cZ_0(z,\bz)  &\equiv & \cD_0^0(z,\bz) = \ln|z| - \ln(1-z) - \ln(1-\bz) \\
    \ln\cZ_L(z,\bz)  &\equiv& \cD_L^0(z,\bz) \\
    \cD_0^k(z,\bz)  & =& \dfrac{\Gamma(k)(z-\bz)^k}{|1-z|^{2k}},\quad k\geq 1 \\
    \bottomrule
    \end{tabular}
    \caption{Summary of definitions}
    \label{tab:my_label_1}
\end{table}

\begin{table}[h!]
    \centering
    \renewcommand{\arraystretch}{1.3}
    \begin{tabular}{>{$}l<{$} >{$}l<{$} >{$}l<{$}}
    \toprule
    \hat{\bf D}_z & =& \tfrac{1}{\beta^2}\tfrac{\partial}{\partial m^2}
      = \tfrac{1}{2\ln|z|}(z\partial_z+\bz\partial_\bz) \\[4pt]
    \hat{\,\bf L}_z & = &\tfrac{i}{\beta}\tfrac{\partial}{\partial \mu}
      = (z\partial_z-\bz\partial_\bz) \\[4pt]
    \hat{\,\bf L}_z^{(k)} & = &\hat{\,\bf L}_z - k \tfrac{z+\bz}{z-\bz} \\[4pt]
    [\hat{\,\bf L}_z]^k & =& (z-\bz)^k \Big(\tfrac{1}{z-\bz}\hat{\,\bf L}_z\Big)^k
       = \Big\{ \mathrm{ord}\prod_{n=0}^{k-1} \Big\}
         \Big[\hat{\,\bf L}_z-n\tfrac{z+\bz}{z-\bz}\Big] \\[4pt]
    [\hat{\bf D}_z,\,\hat{\,\bf L}_z] & = &0, \qquad
       [\hat{\,\bf L}_z^{(k)},\,\hat{\bf D}_z] = 0 \\[4pt]
    \check{\!\mathbf{d}}_{z;z'} & = &\int_0^{|z|}\frac{d|z'|}{|z'|}\,2\ln|z'|, 
       \quad [\,\check{\!\mathbf{d}}_{z;z'},\hat{\,\bf L}_{z'}]=0 \\
    \bottomrule
    \end{tabular}
    \caption{Differential operators $\hat{\bf D}_z$, $\hat{\,\bf L}_z$ 
    and integral operator $\check{\!\mathbf{d}}_{z;z'}$.}
    \label{tab:my_label_2}
\end{table}

\begin{table}[h!]
    \centering
    \renewcommand{\arraystretch}{1.3}
    \begin{tabular}{>{$}l<{$} >{$}l<{$} >{$}l<{$}}
    \toprule
    \ln\cZ_L(z,\bz) & =& [-\,\check{\!\mathbf{d}}]_{z;z'}^L * \ln \cZ_0(z',\bz') \\[4pt]
    \hat{\bf D}_z * \ln \cZ_L(z,\bz) & = &\ln \cZ_{L-1}(z,\bz)
       = -\tfrac{1}{\beta}\langle \cO(z,\bz)\rangle_L \\[4pt]
    \hat{\,\bf L}_z * \ln \cZ_L(z,\bz) & = &\langle\cQ(z,\bz)\rangle_L \\[4pt]
    \check{\!\mathbf{d}}_{z;z'} * \ln \cZ_L(z',\bz') & = &-\ln \cZ_{L+1}(z,\bz) \\[4pt]
    \hat{\,\bf L}_z^{(k)} * \cD_0^k(z,\bz) & = &\cD_0^{k+1}(z,\bz) \\[4pt]
    [\hat{\,\bf L}_z]^k * \ln\cZ_0(z,\bz) & = &\cD_0^k(z,\bz) \\
    \bottomrule
    \end{tabular}
    \caption{Actions of $\hat{\bf D}_z$, $\hat{\,\bf L}_z$ 
    and $\check{\!\mathbf{d}}_{z;z'}$.}
    \label{tab:my_label_3}
\end{table}

\section{The thermal representation of conformal ladder integrals}
\label{sec: correspondence}

\subsection{The correspondence}

\begin{figure}[h!]
    \centering
    \includegraphics[width=0.8\linewidth]{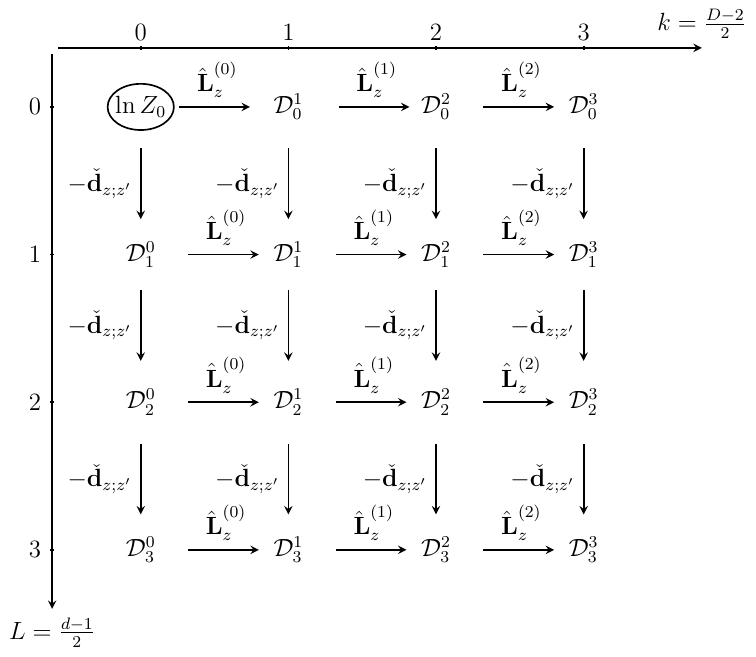}
    \caption{Conformal ladder integrals built from the partition function $\ln \mathcal{Z}_{0}$}
    \label{fig: from HO to conf integrals}
\end{figure}

Comparing our representation (\ref{ILk_our1}) of the conformal ladder integrals with (\ref{Zdexpl}) and (\ref{ZL}) we notice that by the following identifications
\be
\label{identifications}
\ze\leftrightarrow z\,,\,\,\,g^2\leftrightarrow\a^2\,,
\ee
and using the definitions (\ref{ILk_lim}), (\ref{d_def}) and (\ref{ZLd}) we can map the conformal ladder integrals to thermal free energies as 
\begin{mybox}
\begin{align}
\label{ILk_res1}
\cI_L^k(\ze,\bze)=\frac{1}{|1-\ze|^{2k}}\Phi_L^k(\ze,\bze) \longleftrightarrow
&\,\frac{1}{\Gamma(k)L!}\frac{1}{(z-\bz)^k}[-\check{\bf d}_{z;z'}]^L*[\hat{\bf{L}}_{z'}]^k*\ln\cZ_0(z',\bz')\nonumber \, \\
&\equiv\frac{1}{\Gamma(k)L!}\frac{1}{(z-\bz)^k}\cD_L^k(z,\bz)\,.
\end{align}
\end{mybox}
This is one of our main results. It holds for $k\geq 1$ and also serves as the definition of the functions $\cD_L^k(z,\bz)$. An inviting rewriting of (\ref{ILk_res1}) is
\be
\label{ILk_res11}
\Phi_L^k(z,\bz)=\frac{1}{L!}\frac{\cD_L^k(z,\bz)}{\cD_0^k(z,\bz)}\,.
\ee
The limit $k=0$ ($D=2$) of (\ref{ILk_res1}) is discussed in Appendix \ref{App:A} . We see that we can build the conformal integrals $\cI_L^k(z,\bz)$ by the successive application of the operators $\check{\bf d}_{z;z'}$ and $\hat{\,\bf L}_z^{(k)}$ on the free energy $\ln \cZ_0(z,\bz)=\cD_0^0(z,\bz)$ of the parent quantum mechanics model (\ref{BasicHamiltonian}). Our construction is depicted in Fig. \ref{fig: from HO to conf integrals}. Notice that the CFT perturbative condition $g^2\ll1$ maps to the condition $\a^2\ll 1$ which corresponds to low temperature. This seems to indicate some sort of weak/strong coupling duality between $D$-dimensional CFTs and $d$-dimensional thermal field theories which would be exciting to study further.

\subsection{The differential equation}

We are now in a position to show that the functions $\cD_L^k(z,\bz)$ satisfy a remarkable second order differential equation. Consider the representation
\be
\label{lnZL}
\ln \cZ_L(z,\bz)=-\frac{1}{(L-1)!}\int_0^{|z|}\frac{d|z'|}{|z'|}2\ln|z'|(\ln^2|z'|-\ln^2|z|)^{L-1}\ln\cZ_0(z',\bz')\,.
\ee
Since $z$ and $z'$ have the same argument it is useful to write
\be
\label{z'z}
z'=|z'|\left(\frac{z}{\bz}\right)^{\frac{1}{2}}\,,
\ee
when taking the partial derivatives of (\ref{lnZL}) wrt to $z,\bz$. Our aim is to apply the Laplacian (\ref{Laplacian}) to $\ln\cZ_{L}(z,\bz)$. A straightforward calculation yields
\begin{align}
\label{DZL}
-\frac{(L-1)!}{4\b^2}\hat{\bf \Delta}_z*\ln\cZ_L(z,\bz)=& -\frac{1}{2}(L-1)\int_0^{|z|}\frac{d|z'|}{|z'|}2\ln|z'|(\ln^2|z'|-\ln^2|z|)^{L-2}\ln\cZ_0(z',\bz')\nonumber \\
&\hspace{-2cm}+(L-1)(L-2)\ln^2|z|\int_0^{|z|}\frac{d|z'|}{|z'|}2\ln|z'|(\ln^2|z'|-\ln^2|z|)^{L-3}\ln\cZ_0(z',\bz')\nonumber \\
&\hspace{-2cm}-(L-1)\ln|z|\int_0^{|z|}\frac{d|z'|}{|z'|}2\ln|z'|(\ln^2|z'|-\ln^2|z|)^{L-2}[z\partial_z+\bz\partial_\bz]\ln\cZ_0(z',\bz')\nonumber \\
&\hspace{-2cm}+|z|^2\int_0^{|z|}\frac{d|z'|}{|z'|}2\ln|z'|(\ln^2|z'|-\ln^2|z|)^{L-1}\partial_z\partial_\bz\ln\cZ_0(z',\bz')\,.
\end{align}
Firstly we note that 
\be
\label{dZ0}
(z\partial_z+\bz\partial_\bz)\ln\cZ_0(z',\bz')=0\,,
\ee
hence the third line in the r.h.s. of (\ref{DZL}) vanishes. To proceed we use the result
\be
\label{ddZ0}
|z|^2\partial_z\partial_\bz\ln\cZ_0(z',\bz')=-\frac{1}{4}|z'|\partial_{|z'|}|z'|\partial_{|z'|}\ln\cZ_0(z',\bz')\,,
\ee
to integrate by parts the last line in the r.h.s. of (\ref{DZL}). The final result is
\begin{align}
\label{DZ01}
\frac{1}{4\b^2}\hat{\bf \Delta}_z*\ln\cZ_L(z,\bz)=& L(L-1)\frac{1}{(L-1)!}\int_0^{|z|}\frac{d|z'|}{|z'|}2\ln|z'|(\ln^2|z'|-\ln^2|z|)^{L-2}\ln\cZ_0(z',\bz')\nonumber\\
&=-L\ln\cZ_{L-1}(z,\bz)=L\hat{\bf D}_z*\ln\cZ_L(z,\bz)\,,
\end{align}
where in the last equation we have used (\ref{Oaverage}). Noticing then that the Laplacian satisfies
\be
\label{DeltaLz}
[\frac{1}{4\b^2}\hat{\bf \Delta}_z,\hat{\,\bf L}_z^{(k)}]=-2k\frac{|z|^2}{(z-\bz)^2}\hat{\,\bf L}_z^{(1)}\,,
\ee
we can show by induction that $\cD_L^k(z,\bz)$ obeys the following equation
\begin{mybox}
\be
\label{EqDLk}
\left(\frac{1}{4\b^2}\hat{\bf \Delta}_z-L\hat{\bf D}_z+k(k-1)\frac{|z|^2}{(z-\bz)^2}\right)*\cD_L^k(z,\bz)=0\,.
\ee
\end{mybox}
This the second key result of our work. The above equation closely resembles the hyperbolic differential equations satisfied by modular forms and partition functions on the torus \cite{Alessio:2021krn,Aggarwal:2024axv,Dorigoni:2021ngn,Dorigoni:2021jfr,Berg:2019jhh}. In terms of the variables $m,\m$ the equation takes the form
\begin{mybox}
\be
\label{EqDLk1}
\left[\left(\frac{\partial^2}{\partial m^2}+\frac{\partial^2}{\partial\m^2}\right)-\frac{2L}{m}\frac{\partial}{\partial m}-\b^2\frac{k(k-1)}{\sin^2(\b \m)}\right]\cD_L^k(m,\m)=0\,.
\ee
\end{mybox}
For $L=0$ this coincides with the e.o.m. for a scalar field on Euclidean AdS$_2$ (or equivalently the hyperbolic plane ${\cal H}_2$) with metric 
\be
\label{AdS2}
ds^2=\frac{1}{\sin^2(\b\m)}\left(dm^2+d\m^2\right)\,,
\ee
and mass $M^2=\b^2k(k-1)$, with $\b$ associated to the inverse AdS$_2$ radius. This implies that the functions $\cD_0^k(m,\m)$ might be interpreted as bulk-to-boundary propagators in an AdS$_2$/CFT$_1$ setting for operators located at the origin of the boundary and having scaling dimensions $k$. This is reminiscent of a recent observation \cite{Hartnoll:2025hly} on the relation of AdS$_2$/CFT$_1$ to conformal quantum mechanics \cite{Chamon:2011xk}, and it becomes intriguing in view of the connection of the latter to conformal ladder graphs \cite{Isaev:2003tk}.  It would also be very interesting to study whether there exists a similar holographic interpretation for the functions $\cD_L^k(m,\m)$ for $L> 0$.

\subsection{Recap of the correspondence}

The following table shows the relationships between the quantities appearing in the context of conformal integrals and those belonging to the thermal averages of massive scalar theories, which are associated with them via our correspondence.

\begin{table}[h!]
    \centering
    \begin{tabular}{ccc}
    \toprule
    \textbf{Conformal ladder integrals } && \textbf{Thermal one-point functions} \\
    \midrule
    Dimension $D $ &$D=2k+2$& Depth $k$ \\
    Loop order $L$ & $d=2L+1$& Dimension $d$ \\
    Spacetime points $x_{i} = (0,1,z,\infty)$ & $z = e^{-\beta m - i \beta \mu}$ & Mass $m$, chemical potential $\m$ \\
    Dimensionless coupling $g^2$ & $g^2=\a^2$ &Geometric parameter $\alpha^{2} = \dfrac{l^{2}}{4\pi \beta^{2}}$ \\
    \bottomrule
    \end{tabular}
    \caption{Correspondence between conformal ladder integrals and thermal one-point functions.}
    \label{tab:correspondence-CFT}
\end{table}

\section{Applications}
\label{sec: applications}

\subsection{Thermal one-point functions of higher-spin conserved currents}
As discussed above the thermal one-point functions of the operators $\cO$ and $\cQ$ in $d=2L+1$ dimensions are given by conformal ladder integrals in $D=2$ and $D=4$ dimensions respectively
\be
\label{OQ}
\langle\cO(z,\bz)\rangle_L\equiv\beta \cD_{L-1}^0(z,\bz)\,,\,\,\, \langle\cQ(z,\bz)\rangle_L\equiv\cD_L^1(z,\bz)\,.
\ee
These appear in the OPE expansion of the thermal two-point function of a free massive complex scalar field $\phi(x)$ with action (\ref{SE}) given by \cite{Laine:2016hma}   
\be
\label{thermal2pt}
\langle\phi^\dagger(\tau,\vec{x})\phi(0)\rangle^{(L)}\equiv g^{(L)}(\tau,\vec{x};m,\m)=\frac{1}{\b}\sum_{n=-\infty}^\infty\int\frac{d^{d-1}\vec{p}}{(2\pi)^{d-1}}\frac{e^{-i(\omega_n-\m)\tau-i\vec{p}\cdot\vec{x}}}{(\omega_n-\m)^2+\vec{p}^2+m^2}\,,
\ee
where the Matsubara frequencies are $\omega_n=2\pi n/\b$ and again here $d = 2L+1$. The Euclidean coordinates in the thermal geometry
$S^1_{\beta}\times\mathbb{R}^{d-1}$ are $x^\mu=(\tau,\vec{x})$ with period $\tau\sim\tau+\beta$,
$r=|x|$ and  $\theta\in[0,\pi]$ is a polar angle in $\mathbb{R}^{d-1}$. Notice that in the presence of the imaginary chemical potential the scalar fields obey twisted (anyonic) boundary conditions as $\phi(t+\b,\vec{x})=e^{-i\b\m}\phi(\tau,\vec{x})$.

The calculation and the analysis of  (\ref{thermal2pt}) was presented in \cite{Petkou:2018ynm,Petkou:2021zhg,Karydas:2023ufs}. We briefly recap its main points. 
The thermal two-point function in a CFT of a complex scalar $\phi(x)$ with scaling dimension $\Delta_\phi$ in $d=2L+1$ takes the general form\footnote{By conformal symmetry this depends only on $r$ and $\theta$ defined above. The momentum space version of this formula appeared for the first time  in \cite{Petkou:1998fb}.} \cite{Iliesiu:2018fao} 
\begin{equation}
\label{phiphi}
g^{(L)}(r,\theta;0,0)=
\frac{1}{r^{2\Delta_\phi}}\left[C_\phi^L(1)+\sum_{\{{\cO}_s\}}a^L_{\cO_s}\left(\frac{r}{\beta}\right)^{\Delta_{\cO_s}}
C_s^{\nu}(\cos\theta)\right]\, ,
\end{equation}
where $\nu = L-1/2$ and $C_{s}^{\nu}(\cos\theta)$ are Gegenbauer
polynomials. To arrive in (\ref{phiphi}) one assumes the existence of a conformal OPE at zero temperature such that $\phi^\dagger\times\phi$ can be expanded in a sum of 
quasiprimary operators ${\cO}_s$ with definite spin $s$ and scaling dimension
$\Delta_{\cO_s}$. The latter are represented by symmetric, traceless and conserved rank-$s$ tensors, and their one-point functions depend on a single parameter which is proportional to the coefficients $a^L_{\cO_s}$. 

For massless free complex scalars when $\Delta_\phi=L-1/2$ one obtains \cite{Petkou:1998fb,Iliesiu:2018fao,Petkou:2018ynm}
\begin{equation}
\label{azeta}
a^L_{\cO_s}=2C_\phi^L(1)\zeta(2L-1+s)\,,\,s=0,2,4...\,.
\end{equation}
In that case, only conserved higher-spin operators with dimensions \mbox{$\Delta_{\cO_s}=2L-1+s$} and even spin $s$ appear in (\ref{phiphi}). Then, all but the leading term in the expansion (\ref{phiphi}) - which is the contribution of the unit operator and corresponds to the zero temperature result - are of the form $r^s C_s^{L-1/2}(\cos\theta)$ and are annihilated by the $d$-dimensional Laplacian $\Box_d$. Then, the name of the game in the thermal bootstrap\footnote{An important activity towards understanding thermal CFT correlators has recently emerged. A recent review with the relevant literature is \cite{Miscioscia:2025pjh}.} is  to understand how interactions alter the operator spectrum and the parameters $\Delta_{\cO_s}$ and $a^{L}_{\cO_s}$ in (\ref{phiphi}). 

In \cite{Petkou:2018ynm,Petkou:2021zhg,Karydas:2023ufs} it was observed that even for massive free field theories such as (\ref{SE}) a part of the thermal two-point function can be expanded in an infinite series of thermal one-point functions of higher-spin conserved currents. This part, which we call {\it the integrable part} of the thermal two-point function, consists of terms of the form $r^s C_s^{L-1/2}(\cos\theta)$ and it is still annihilated by the $d$-dimensional Laplacian $\Box_d$. We then notice that the mass and imaginary chemical potential relevant deformations just change the functional form of the thermal one-point function coefficients $a^L_{\cO_s}$ of the integrable part of the thermal two-point function. However, they do induce a non-trivial effect in the form of an infinite tower of ``shadow" operators in the spectrum, as it was discussed in \cite{Petkou:2018ynm}. The explicit calculations give (see also \cite{Kumar:2025txh})
\begin{align}
\label{GLgL}
g^{(L)}(r,\theta;m,\m)&=
\frac{1}{(2\pi)^{\nu}}\sum_{n=-\infty}^\infty\!e^{i\mu n}\left[\frac{m}{|X_n|}\right]^{\nu}\!\!K_{\nu}(m|X_n|)\nonumber \\
&=\frac{1}{r^{2\Delta_\phi}}\left[1+\sum_{\{{\cO}_s\}}a^L_{\cO_s}(z,\bz)\left(\frac{r}{\beta}\right)^{\Delta_{\cO_s}}
C_s^{\nu}(\cos\theta)+\text{``shadows"}\right]
\end{align}
with $X_n=(\tau-n,\vec{x})$ and $K_\nu(x)$ the modified Bessel functions of the second kind.
The new coefficients $a^L_{\cO_s}(z,\bz)$, where $z,\bz$ are the previously defined modular variables, can be calculated from (\ref{GLgL}) using the inversion method of \cite{Iliesiu:2018fao}, taking into account  that the two-point function is complex so that the discontinuities along the cuts in the positive and negative $r$-axis are complex conjugates. The result is \cite{Karydas:2023ufs},
\begin{align}
    \label{aOs}
    \hspace{-1cm}a^{L}_{\cO_s}(z,\bz)&=\frac{\Gamma\left(L-\frac{1}{2}\right)}{\Gamma\left(L+s-\frac{1}{2}\right)(4\pi)^L 2^{2s}}\sum_{n=0}^{L-1+s}\frac{2^n}{n!}\frac{(\beta m)^n(2L-2+s-n)!}{(L-1+s-n)!}\nonumber \\
    &\hspace{3cm}\times\left[Li_{2L-1+s-n}(z)+(-1)^s Li_{2L-1+s-n}(\bar{z})\right]\,,
\end{align}
and by virtue of (\ref{OQ}) we have
\begin{equation}
    \label{aOaQ}
    a^L_{\cO_0}(z,\bz)=\frac{1}{(4\pi)^L\alpha^{2L}}\cD_{L-1}^{0}(z,\bz)\,,\,\,a^L_{\cO_1}(z,\bz)=\frac{1}{(4\pi)^L\alpha^{2L}}\frac{1}{2}\cD_L^{1}(z,\bz)\,,
\end{equation}
 
Notice now that  $a^L_{\cO_s}(z,\bz)$ would correspond to thermal one-point functions of symmetric, traceless and conserved tensors, which however are {\it not} the operators of the massive theory (\ref{SE}) e.g. even the energy momentum tensor of the latter is not traceless. Nevertheless, the fact that the theory is free allows us to consistently construct the higher-spin currents whose thermal contributions appear in (\ref{GLgL}), see an explicit example in \cite{Karydas:2023ufs}.

It is then natural to ask if the higher-spin thermal one-point functions are related to the conformal ladder integrals. The answer was presented in \cite{Karydas:2023ufs} where by a brute force calculation we derived the following recursive relations
\begin{equation}
    \label{recursion}
    a^L_{\cO_{s+2}}(z,\bz)=\frac{2\pi}{2L-1}a^{L+1}_{\cO_s}(z,\bz)+\frac{\ln^2|z|}{(2L-1+2s)(2L+1+2s)}a^L_{\cO_s}(z,\bz)\,,
\end{equation}
valid for $s\geq 0$. This demonstrates that the thermal one-point functions of the spin-zero and spin-one operators determine the thermal one-point functions of all higher-spin currents. The proof of this relationship is rather technical and involves using the equations of motion of the free theory. We present it in detail in Appendix \ref{App:D} as we believe that it provides the setup for studying thermal one-point functions of Wilson-Fisher-like critical theories whose fields obey non-trivial equations of motion. 

\begin{figure}[h!]
    \centering
    \includegraphics[width=0.8\linewidth]{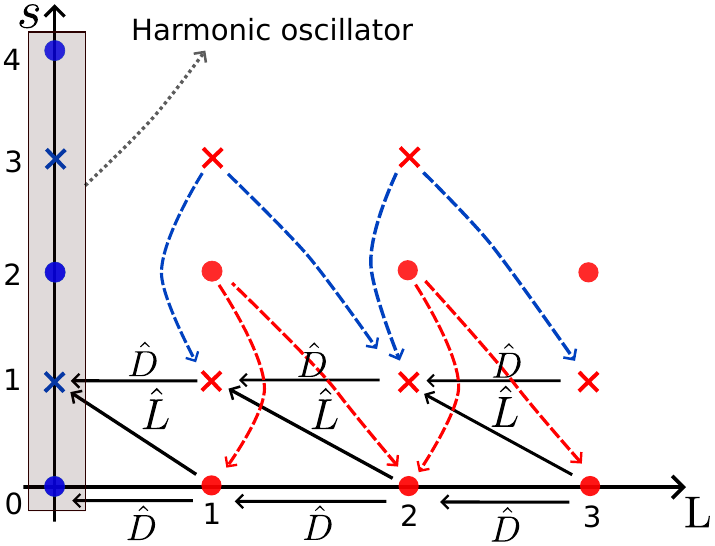}
    \caption{Relationships between the $a_{\mathcal{O}_{s}}$, the dashed lines represent an algebraic relationship while the plain lines stand for a differential one.}
    \label{fig: relations aL}
\end{figure}

The relationships between the various coefficients $a^{L}_{\cO_{s+2}}$ are depicted in Fig. \ref{fig: relations aL}.
Equivalently, (\ref{recursion}) shows that the {\it integral part} of the thermal two-point function is determined only by conformal ladder integrals in $D=2$ and $D=4$. Hence, the higher-spin thermal one-point functions in $d=2L+1$ dimensions correspond to linear combinations of multi-loop ladder integrals in $D=2$ (for even spin) and $D=4$ (for odd spin). We can think of (\ref{recursion}) as a generalization, for both even and odd spins $s$, of the trivial correspondence exhibited in (\ref{aOaQ}) between the thermal one-point functions of spin-$s$ conserved currents in $d=2L+1$ and spin-$s-2$ conserved currents in $d=2L+3$ dimensions.

\subsection{The hyper-partition function and the all-loop resummation}

The representation (\ref{ZL}) motivates us to consider a {\it hyper-partition function} and the corresponding {\it hyper  free energy} as
\be
\label{hyperZ}
\cZ(z,\bz;\a^2)=\prod_{L=0}^\infty \cZ_L(z,\bz;\a^2)\,,\,\,\,\,\ln\cZ(z,\bz;\a^2)=\sum_{L=0}^\infty \ln\cZ_L(z,\bz;\a^2)\,.
\ee
It may appear bizarre to consider a partition function for thermal systems across their dimension $d$, and even to consider the limit of infinite dimensions, but as we have seen before $d$ is mapped to the loop order $L$ of conformal integrals, which can take any integer value and be sent to infinity. By virtue of e.g. (\ref{ZL}) and (\ref{Oaverage})  we can derive the following first order equation for the {\it hyper-free energy}
\be
\label{dhyperF}
(\hat{\bf D}_z+\a^2)*\ln\cZ(z,\bz;\a^2)+\ln\cZ_{-1}(z,\bz)=0\,,
\ee
which in the variables $m,\m$  takes the form 
\be
\label{dhyperF1}
\left(\frac{\partial}{\partial m}+2\a^2\b^2 m\right)\ln\cZ(m,\m;\a^2)=-\b\frac{\sinh\b m}{\cosh\b m-\cos\b\m}\,.
\ee
A quick way to evaluate $\ln\cZ(m,\m;\a^2)$ is directly from the first line of  (\ref{Zdexpl}) where the sum can be performed leading to 
\begin{align}
\label{dhyperF2}
\ln\cZ(m,\m;\a^2)&=\ln\cZ_0(m,\m)+\sum_{L=1}^\infty \ln\cZ(m,\m;\a^2)\nonumber \\
&=\ln\cZ_0(m,\m)+\a^2\b^2 e^{-\a^2\b^2 m^2}\int_m^\infty 2\omega \,d\omega \,e^{\a^2\b^2 \omega^2}\ln\cZ_0(m,\m)\,.
\end{align}
An integration by parts gives
\begin{align}
\label{dhyperF3}
\ln\cZ(m,\m;\a^2)&=\ln\cZ_0(m,\m)+e^{-\a^2\b^2 m^2}\left[e^{\a^2\b^2 \omega^2}\ln\cZ_0(\omega,\m)\right]_m^\infty \nonumber\\
&+\b\,e^{-\a^2\b^2 m^2}\int_m^\infty d\omega \,e^{\a^2\b^2 \omega^2}\frac{\sinh(\b\omega)}{\cosh(\b\omega)-\cos(\b\m)}\,.
\end{align}
The $\ln\cZ_0(m,\m)$ cancel in the first line of (\ref{dhyperF3}), and we remain with a divergent contribution at infinity, and a divergent integral. Interestingly, the analytic continuation $\a\mapsto i\alpha$ renders the result finite. This is relevant for a similar calculation of the all-loop resummation of ladder integrals that can be easily done using our formalism. Indeed, from (\ref{ILk_res1}) we have
\be
\label{sum1}
\cI^k(z,\bz;g^2)=\sum_{L=0}^\infty (-g^2)^{L}\cI_L^k(z,\bz)=\frac{1}{\Gamma(k)(z-\bz)^k}\left[\hat{\,\bf L}_z\right]^k*\sum_{L=0}^\infty\frac{1}{L!}\ln\cZ_L(z,\bz;-g^2)\,,
\ee
and using that we can obtain the all-loop sum of conformal ladder integrals (\ref{ILk_def}) for any $D=2k+2$ by acting with $\left[\hat{\,\bf L}_z\right]^k$ on the {\it Borel sum} of the free energies $\ln\cZ_L(z,\bz;-g^2)$. The latter can be acquired by first calculating from the  first line of (\ref{Zdexpl}) getting
\be
\label{dhyperF4}
\sum_{L=0}^\infty \frac{1}{L!}\ln\cZ_L(m,\m;\a^2)=\ln\cZ_0(m,\m)+\a\b\int_m^\infty 2\omega d\omega\, \frac{I_1(2\a\b\sqrt{\omega^2-m^2})}{\sqrt{\omega^2-m^2}}\ln\cZ_0(m,\m)\,.
\ee
where $I_k(z)$ is the modified Bessel function of the first kind\footnote{Explicitly we have $\sum_{n=1}^\infty\frac{z^{2(n-1)}}{n!(n-1)!}=\frac{1}{z}I_1(2z)$.} and then do an analytic continuation $\a\to ig$.  Using the standard formula $I_1(z)=\frac{d}{dz}I_0(z)$ we can integrate by parts to find
\begin{align}
\label{dhyperfF5}
\sum_{L=0}^{\infty}\frac{1}{L!}\ln\cZ_L(z,\bz;\a^2)&= \ln\cZ_0(m,\m) +\left[I_0(2\a\b\sqrt{\omega^2-m^2})\ln\cZ_0(\omega,\m)\right]_m^\infty\nonumber\\
&+\b\int_m^\infty d\omega\,I_0(2\a\b\sqrt{\omega^2-m^2})\frac{\sinh(\b\omega)}{\cosh(\b\omega)-\cos(\b\m)}\,.
\end{align}
Again we notice that $\ln\cZ_0(m,\m)$ cancel in the first line of (\ref{dhyperfF5}), and we remain with a divergent contribution at infinity, and a divergent integral. Doing the analytic continuation $\a\to ig$, we find $I_0(2\a\b\sqrt{\omega^2-m^2})$ will be replaced by
\be
I_0(2ig\b\sqrt{\omega^2-m^2}) = J_0(2g\b\sqrt{\omega^2-m^2})\,
\ee
where $J_0$ is the zeroth order Bessel function of the first kind. In the end, we find
\begin{align}
\label{dhyperfF6}
\sum_{L=0}^{\infty}\frac{1}{L!}\ln\cZ_L(z,\bz;-g^2)&= \ln\cZ_0(m,\m) +\left[J_0(2g\b\sqrt{\omega^2-m^2})\ln\cZ_0(\omega,\m)\right]_m^\infty\nonumber\\
&+\b\int_m^\infty d\omega\,J_0(2g\b\sqrt{\omega^2-m^2})\frac{\sinh(\b\omega)}{\cosh(\b\omega)-\cos(\b\m)}\,.
\end{align}
Note that when $\omega\to\infty$, $\ln Z_0(\omega,\m)$ contains only the zero energy contribution $-\b\omega$ and thus the first line of \eqref{dhyperfF6} does not depend on $\m$. Therefore, to get $\cI^k(z,\bz;g^2)$ for $k\geqslant1$, ($D\geq 4$), one only needs to act with $\left[\hat{\,\bf L}_z\right]^k$ on the second line of \eqref{dhyperfF6}. For example, when $k=1$, we have 
\begin{align}
\cI^k(z,\bz;g^2) &= \frac{i}{z-\bz} \frac{\partial}{\partial\m}\left[\int_m^\infty d\omega\,J_0(2g\b\sqrt{\omega^2-m^2})\frac{\sinh(\b\omega)}{\cosh(\b\omega)-\cos(\b\m)}\right]\nonumber\\
&=\frac{1}{2|z|}\int_m^\infty d\omega\,J_0(2g\b\sqrt{\omega^2-m^2})\frac{\sinh(\b\omega)}{\left(\cosh(\b\omega)-\cos(\b\m)\right)^2}\,,
\end{align}
which matches with the all-loop resummation by Broadhurst \cite{Broadhurst:1993ib,Broadhurst:2010ds}, see also \cite{Giombi:2020enj,Caetano:2023zwe,Brown:2025cbz}.

\section{Summary and Outlook}
\label{sec: conclusion}

In this work we have established a {\it thermal representation of conformal ladder integrals.} This is an explicit relationship between conformal ladder integrals in even $D$ dimensions for any loop order $L$, and thermal averages in free massive theories of complex scalars. The relationship associates the loop order $L$ of the integrals to the dimension on which the thermal theory resides as $d=2L+1$, and the dimension $D$ of the integrals to a {\it depth} parameter $k$ of the thermal theory as $D=2k+2$. Notably, all integrals and their corresponding thermal averages can be constructed from a {\it parent quantum mechanical system} that consists of a set of harmonic oscillators twisted by an imaginary chemical potential. Using our thermal representation we show that all conformal ladder integrals satisfy a familiar second order differential equation. Applications to thermal two-point functions and to the all-loop resummation of ladder integrals are also discussed.

At first sight our thermal representation appears to be a mathematical map between very different physical quantities, such as the map of spacetime cross-ratios $\ze$ in (\ref{ILk_lim}) to the complex exponentials of a thermal field theory $z$ in (\ref{z}). Nevertheless, we believe that it offers some novel perspectives in a number of research directions.
\begin{itemize}

\item
Although we do not have at the moment a deeper understanding of our representation we cannot help noticing the similarity with the AGT conjecture \cite{Alday:2009aq} and the huge amount of related work (see \cite{LeFloch:2020uop}) for a recent review). In the AGT case, however, it was possible to view the spacetime cross-ratios as moduli on complex structures which makes their identification as instanton counting parameters more clear. In our case, however, it is not immediately clear how to use the full power of supersymmetry and complex geometry, since we do not rely on them to derive our results. 

\item
The all-loop resummation of conformal ladder integrals have recently appeared in calculations of four-point correlators in large-charge sectors of the $O(N)$ vector model \cite{Giombi:2020enj} and ${\cal N}=4$ SYM \cite{Caetano:2023zwe,Brown:2025cbz}. Related works include \cite{Coronado:2018ypq,Aprile:2025hlt,Aprile:2024lwy}. Our thermal representation offers a novel perspective for correlators in large-charge sectors as we associate them with a sum of free energies that is reminiscent of the geometric/genus expansion of string amplitudes.

\item
Thermal one-point functions in CFTs on manifolds of the form $S^1\times\mathbb{R}^{d-1}$ have been recently studied for zero and non-zero chemical potential in a number of works including \cite{David:2023uya,David:2024naf, Kumar:2025txh, Diatlyk:2023msc, Benedetti:2023pbt}. From our perspective the results presented there correspond to combinations of conformal integrals in $D=2$ and $D=4$ dimensions and it would be interesting to verify this. It would also be interesting to study how our differential equation and recursive relations are modified away from the gaussian fixed points. Finally, we expect that the conformal integrals for $D\geq 6$ will appear in thermal two-point functions of composite operators and it would be nice to study this further. 

\item
One-point functions in CFTs on manifolds of the form $S^1\times S^{d-1}$ have also attracted attention in recent studies such as \cite{David:2024pir,Alkalaev:2024jxh,Buric:2024kxo,Buric:2025uqt}. These are more complicated than the ones on the thermal geometry and are expanded in the so-called thermal conformal blocks \cite{Gobeil:2018fzy}. Further, it was pointed out in \cite{Alkalaev:2024jxh} that scalar thermal conformal blocks are closely related to conformal integrals too. Since thermal one-point functions arise as large volume limits of the one-point functions on $S^1\times S^{d-1}$  it would be interesting to study how the differential equations and the other relations of the conformal ladder integrals arise from the thermal conformal blocks. 

\item
Some interesting studies of thermal two-point functions in the context of the thermal conformal bootstrap have recently appeared \cite{Marchetto:2023xap,Barrat:2025wbi,Barrat:2025nvu,Buric:2025anb,Buric:2025uqt}. A crucial issue here is the study of the constraints imposed on thermal one-point functions by the KMS condition. As thermal one-point functions are related to conformal integrals, it is conceivable that various symmetry property of the latter are related to the KMS condition. It would be interesting to study this further.

\item
A number of recent works have discussed the modular properties of thermal partition functions in CFTs \cite{Shaghoulian:2015kta,Shaghoulian:2016gol,Alessio:2021krn,Aggarwal:2024axv,Allameh:2024qqp}. From our perspective, the modular properties of the partition functions should have a counterpart in the conformal integrals. Indeed, the relevant formulae in terms of polylogarithms are very similar \cite{Alessio:2021krn,Aggarwal:2024axv} and furthermore the relevant free energies satisfy differential equations and recursive relations similar to the ones presented in this work (see  \cite{Alessio:2021krn}). Such observations open up the possibility of connecting the mathematics of modular forms that plays a prominent role in studies of string amplitudes \cite{Dorigoni:2021ngn,Dorigoni:2021jfr,Berg:2019jhh}, to thermal CFTs. 

\item
An interesting class of conformal theories are those described by higher derivative actions e.g. \cite{Brust:2016gjy}. Conformal ladder integrals in such theories have been recently studied in \cite{Giombi:2022gjj} and their thermal properties in \cite{Benedetti:2023pbt}. We think that our thermal representation can be applied in the study of these models which are thought to describe long-range critical systems. 

\item
Our thermal representation of conformal ladder integrals may also provide a useful new perspective to the numerous studies on  the integrability and other mathematical properties properties of multiloop Feynman graphs.\footnote{It is hard to do justice to the wealth of important works in this direction which include among others \cite{Basso:2017jwq,Vanhove:2018elu, Charlton:2021uhu,Travaglini:2022uwo}.} Recent examples are the use our our $\hat{\bf D}_z$ and $\hat{\,\bf L}_z$ operators in \cite{Loebbert:2024fsj} to show that ladder integrals obey Toda-like equations, and the proof of antipodal symmetry in \cite{Dixon:2025zwj} using some of our first order differential equations. One may also try to extend the thermal representation to pentaladders \cite{Caron-Huot:2018dsv}. We believe that there is more to be uncovered in these directions see e.g. \cite{He:2025lzd} for a recent work.. 

\item
Finally, it is worth pointing out the versatility of our parent quantum mechanical system (\ref{BasicHamiltonian}) which has been recently appeared in the form of a {\it primon gas} whose partition function corresponds to Wheeler-DeWitt eigenfunctions that describe quantum gravity near a singularity \cite{Hartnoll:2025hly,DeClerck:2025mem}. Intriguingly, the latter satisfy differential equations similar to our (\ref{EqDLk1}) and they are intriguingly related to AdS$_2$/CFT$_1$ through conformal quantum mechanics \cite{Chamon:2011xk}. 

\end{itemize}

\acknowledgments

A.C.P. wishes to acknowledge numerous enlightening discussions and correspondence at various stages of this work with  G. Barnich, M. Berg, J. David, S. Giombi, V. Kazakov, A. Kleinschmidt, E. Marchetto, A. Miscioscia, N. Obers, E. Pomoni, A. Santambrogio, L. Shumilov, C. Wen and K. Zarembo. He also thanks the Simons Center for
Geometry and Physics at Stony Brook for its hospitality during the program `Black
Hole Physics from Strongly Coupled Thermal Dynamics’, where part of this work
was presented.
The work of M.V. was supported by the Fonds de la Recherche Scientifique F.R.S. –
FNRS under the Grant No. T.0047.24. A.C.P. and M.V. would like to acknowledge the \emph{Hubert Curien Franco-Greek partnership} for partially financing scientific visits to each other institutions during the completion of this work. 
\appendix

\section{Conformal ladder integrals in two dimensions}
\label{App:A}

One way to define the $L$-loop conformal ladder integrals is as the leading $N_c$ contributions in the calculation of the four-point function 
\begin{equation}
G^{(L)}_{D,\omega}(\{x_i\})=\langle\Tr\left[\phi_2^L(x_1)\phi_1(x_3)\phi_2^{\dagger L}(x_2)\phi_1^{\dagger}(x_4)\right]\rangle\,,
\end{equation}
for the  bi-scalar theory in $D$-dimensions introduced in \cite{Kazakov:2018qbr} with Lagrangian
\be
\label{fishnet}
    {\cal L}=N_c\Tr\left[\phi_1^\dagger(-\partial^2)^{\omega}\phi_1+\phi_2^\dagger(-\partial^2)^{\frac{D-2\omega}{2}}\phi_2+a_{D,\omega}^2\phi_1^\dagger\phi_2^\dagger\phi_1\phi_2\right]\,,
\ee
where the scalar fields $\phi_{1,2}$ belong to the adjoint of $SU(N_c)$, $\omega\in\left(0,\frac{D}{2}\right)$ and the coupling $a_{D,\omega}^2$ is classically dimensionless. 
For $D=2,\omega=1$ the model (\ref{fishnet}) is singular and $G^{(L)}_{2,1}$ would appear to vanish. Nevertheless, a nonzero result can be obtained if we define the effective coupling \cite{Karydas:2023ufs}
\begin{equation}
    \label{a21}
    \tilde{a}_{D,\omega}=a_{D,\omega}\frac{1}{\Gamma(D/2-\omega)}\,,
\end{equation} 
that remains finite as $D\mapsto 2, \omega\mapsto 1$. Then, following the graph-building techniques introduced in \cite{Derkachov:2018rot,Olivucci:2021cfy,Derkachov:2021ufp,Derkachov:2023xqq} one can show that the appropriately normalised four-point function of Fig. \ref{fig: conf integrals}  is given by
\begin{equation}
\label{GL21}
    \tilde{G}^{(L)}_{2,1}(z,\bz)=\tilde{a}_{2,1}^{2L}\sum_{m\in\mathbb{Z}}\int d\nu \frac{(z\Bar{z})^{i\nu}(z/\Bar{z})^{m/2}}{(\frac{m^2}{4}+\nu^2)^{L+1}}\,.
\end{equation}
Since $|z|<1$ we compute the integrals above using contour intergation. When $m\neq 0$ we close the contour from below and pick up the residues in the lower half complex plane. We obtain
\be
\label{GL21m}
   \sum_{m\neq 0}\int d\nu \frac{(z\Bar{z})^{i\nu}(z/\Bar{z})^{m/2}}{(\frac{m^2}{4}+\nu^2)^{L+1}}=  \frac{2\pi}{L!}\sum_{n=0}^{L} \frac{ (2L-n)! (-2\ln |z|)^n}{(L-n)!n!} 2\Re[Li_{2L+1-n}(z)].
\ee
For $m=0$ the contour integral appears to be zero, but there is a pole on the real axis. Taking the Cauchy principal value we obtain 
\begin{equation}
    -\int_{C_\epsilon} d\nu \frac{|z|^{2i\nu}}{\nu^{2L+2}}=-i\int_\pi^{2\pi} d\theta \frac{\exp{(2i\epsilon\ln|z| e^{i\theta})}}{\epsilon^{2L+1}e^{i(2L+1)\theta}}\,.
\end{equation}
For $\epsilon\mapsto 0$ we encounter $2L+1$ divergent terms, which we discard, and a finite contribution which reads
\begin{equation}
\label{GL210}
    -i\int_\pi^{2\pi} d\theta \frac{(2i\ln|z|)^{2L+1}}{(2L+1)!}=(-)^L\pi\frac{(2\ln|z|)^{2L+1}}{(2L+1)!}\,.
\end{equation}
Putting together (\ref{GL21m}) and (\ref{GL210}) we finally obtain
\begin{equation}
\label{GL21Zl}
    G^{(L)}_{2,1}(z,\bz)\equiv \cI_L^0(z,\bz)=\frac{2\pi}{L!}\ln \cZ_L(z,\bz)\,,
\end{equation}
which is the $k=0$, $D=2$, limit of (\ref{ILk_res1}). 

\section{Partition functions of free massive complex scalars in $d=2L+1$ dimensions}
\label{App:B}
In the standard imaginary time formalism the theory lives on $S^1_\b\times{\mathbb R}^{d-1}$ with Euclidean coordinates $x^\m=(\tau,\bar{x})$, $\tau\in [0,\b]$ and the scalar fields $\phi(x)$ obey periodic boundary conditions $\phi(t+\b,\vec{x})=\phi(\tau,\vec{x})$. To evaluate the path integral for  complex scalars in odd $d$-dimensions with action (\ref{SE}) 
one removes the short distance singularity by subtracting the zero temperature, mass and chemical potential results as in (\ref{Zgcphi}). Standard manipulations lead to 
\begin{align}
\label{eq:A1}
  \frac{1}{V_L}\ln \cZ_L &= -\frac{1}{\beta}\sum_{n=-\infty}^\infty\int\frac{d^{2L}\Vec{p}}{(2\pi)^{2L}}\ln\left[(\omega_n-\mu)^2+\Vec{p}^2+m^2\right] + \int\frac{d^{2L+1}p}{(2\pi)^{2L+1}}\ln\left(p^2\right) \nonumber\\
    &= \ln \cZ_L^{\text{th}} + \ln \cZ_L^{\text{z-t}}\,,
\end{align}
where  the Matsubara frequncies are $\omega_n=2\pi n/\b$ and we consider the space to be a very large box with sides of length $l$, such that the spacetime volume $V_L\sim \beta l^{2L}$. In (\ref{eq:A1}) the thermal and zero-temperature contributions are 
\begin{align}
    \ln \cZ_L^{\text{th}} &= -\frac{1}{\beta}\sum_{n=\infty}^\infty\int\frac{d^{2L}\Vec{p}}{(2\pi)^{2L}}\ln\left[(\omega_n-\mu)^2+\Vec{p}^2+m^2\right] + \int\frac{d^{2L+1}p}{(2\pi)^{2L+1}}\ln\left(p^2+m^2\right)\,,\label{logzthermal}
 \\
\ln \cZ_L^{\text{z-t}} &= -\int\frac{d^{2L+1}p}{(2\pi)^{2L+1}}\ln\left(\frac{p^2+m^2}{p^2}\right)\,.
\end{align}
Shifting the variable $p_0\to p_0-\mu$ the second term of \eqref{logzthermal} becomes
\begin{equation}
    \int \frac{dp_0}{2\pi} \int\frac{d^{2L}\Vec{p}}{(2\pi)^{2L}}\ln\left[(p_0-\mu)^2+\Vec{p}^2+m^2\right]\,.
\end{equation}
We can then apply eq. (2.34) of \cite{Laine:2016hma} and find (defining $\omega_{\Vec{p}} = \sqrt{\Vec{p}^2+m^2}$)
\begin{equation}
    \ln \cZ_L^{\text{th}} = -\int\frac{d^{2L}\Vec{p}}{(2\pi)^{2L}}\int_{-\infty-i0^+}^{+\infty-i0^+} \frac{dp_0}{2\pi} \frac{\ln\left[(p_0-\mu)^2+\omega_{\Vec{p}}^2\right]+\ln\left[(p_0+\mu)^2+\omega_{\Vec{p}}^2\right]}{e^{i\beta p_0}-1}\,.
\end{equation}
Performing the change of variables $p_0\to p_0\pm\mu$ for the two terms respectively followed by an integration by parts, we find
\begin{equation}
    \ln \cZ_L^{\text{th}} = \int\frac{d^{2L}\Vec{p}}{(2\pi)^{2L}}\int_{-\infty-i0^+}^{+\infty-i0^+} \frac{dp_0}{2\pi i\beta}\frac{2p_0}{p_0^2+\omega_{\Vec{p}}^2}\left[\ln(1-e^{-i\beta p_0-i\beta\mu})+\ln(1-e^{-i\beta p_0+i\beta\mu})\right]\,.
\end{equation}
Closing the $p_0$ contour in the lower half  plane  to pick the pole at $p_0 = -i\omega_{\Vec{p}}$ we find
\begin{eqnarray}
    \ln \cZ_L^{\text{th}} 
    &=& -\frac{1}{\beta}\int\frac{d^{2L}\Vec{p}}{(2\pi)^{2L}}2\Re\left(\ln(1-e^{-\beta\sqrt{\Vec{p}^2+m^2} -i\beta\mu})\right)\nonumber\\
    &=& -\frac{S_{2L}}{(2\pi)^{2L}\beta}2\Re\left(\int_m^\infty d\omega \omega (\omega^2-m^2)^{L-1} \ln(1-e^{-\beta\omega -i\beta\mu})\right)\,.
\end{eqnarray}
We also have
\begin{eqnarray}
    \ln \cZ_L^{\text{z-t}} =-\int\frac{d^{2L+1}p}{(2\pi)^{2L+1}}\ln\left(\frac{p^2+m^2}{p^2}\right)=-\frac{m^{2L+1}S_{2L+1}}{(2\pi)^{2L+1}}\int_0^\infty dx x^{2L}\ln \left(1+\frac{1}{x^2}\right)\,,
\end{eqnarray}
where $x^2=p^2/m^2$.  We will evaluate the integral for $0<L<\frac{1}{2}$ where it converges and then do an analytic continuation. Integrating by parts (or looking at \cite{Gradshteyn:1943cpj}) we find
\begin{eqnarray}
    \int_0^\infty dx x^{2L}\ln \left(1+\frac{1}{x^2}\right) =\frac{2}{2L+1} \int_0^\infty dx \frac{x^{2L}}{1+x^2}
  =\frac{\pi}{(2L+1)\sin(\pi(L+\frac{1}{2}))}\,.
\end{eqnarray}
This formula can be continued for all $L$,  and in particular for integer $L$ we have
\begin{equation}
    \ln \cZ_L^{\text{z-t}} = \frac{(-1)^{L+1}m^{2L+1}S_{2L+1}}{2(2L+1)(2\pi)^{2L}}\,.
\end{equation}
Putting all together we obtain
\begin{equation}
    \label{eq: free energy}
 \frac{1}{V_L} \ln\cZ_L =- \frac{S_{2L}}{(2\pi)^{2L}\beta}2\Re\int_m^\infty d\omega \omega (\omega^2-m^2)^{L-1} \ln(1-e^{-\beta\omega -i\beta\mu}) - \frac{(-1)^{L}m^{2L+1}S_{2L+1}}{2(2L+1)(2\pi)^{2L}}\, .
\end{equation}
The above formula is the regularized version of (\ref{Zdexpl}) as the integral is now convergent. It has been performed in \cite{Petkou:2021zhg} and the result yields exactly (\ref{ZLresult}).

\section{Expansion in Zagier-Ramakrishnan single-valued polylogarithms}
\label{App:E}

The single-valued Zagier-Ramakrishnan polylogarithms are defined as \cite{Zagier2,Goncharov:1996dce}
\begin{equation}
\label{C1}
    \mathcal{L}_n(z) = \mathcal{R}_n \left[\sum_{k=0}^{n-1} \frac{B_k}{k!}(2\ln|z|)^k Li_{n-k}(z)\right]\,,
\end{equation}
where $\mathcal{R}_n$ is taking real or imaginary part for $n$ odd and even respectively. The $B_k$'s are the Bernoulli numbers. We will show here how $\langle \cQ(z,\bz)\rangle_L=\hat{\,\bf L}_z*\ln\cZ_L(z,\bz)$, for $L\geqslant1$, is expanded in terms of the $ \mathcal{L}_n(z)$. From (\ref{ZLresult}) and (\ref{Qaverage}) we obtain
\begin{equation}
\label{C2}
    \langle \cQ(z,\bz) \rangle_L = \sum_{n=0}^L \frac{(-)^n(2L-n)!}{(L-n)!n!} (2\ln|z|)^{n} (Li_{2L-n}(z)-Li_{2L-n}(\Bar{z}))\,.
\end{equation}
Since the above contains only the imaginary part of the polylogarithms we make the following ansatz
\begin{equation}
\label{C3}
    \langle \cQ \rangle_L = 2i(\alpha^2)^L \sum_{j=0}^{j^{(L)}_{\max}} c^{(L)}_j (2\ln|z|)^{2j} \mathcal{L}_{2L-2j}(z)\,,
\end{equation}
for some coefficients $c^{(L)}_j$. Notice that the Bernoulli number $B_k$ vanishes for $k$ being any odd number larger than one, and thus $\Im(Li_{2l+1})$ only appears in $\mathcal{L}_{2l+2}$. As $\langle \cQ(z,\bz) \rangle_L$ contains a sum of $\Im(Li_i)$ with $L\leq i\leq 2L$, this implies that $\mathcal{L}_{2L-2j}$ with $2L-2j-1<L$ does not appear in (\ref{C3}). Thus $j^{(L)}_{\max} = \lceil\frac{L}{2}\rceil-1$, where $\lceil\frac{L}{2}\rceil$ denotes the smallest integer that is greater than or equal to $\frac{L}{2}$.

By comparing the coefficients in front $\Im(Li_{2L-2j-1})$ we then find
\begin{equation}
    c^{(L)}_j = \frac{2(2L-2j-1)!}{(L-2j-1)!(2j+1)!}\,.
\end{equation}
What remains to be checked is whether the coefficients in front of $\Im(Li_{2L-2l})$ for  $0\leq l\leq L-1$ from $0$ match  the ones in $\langle \cQ(z,\bz) \rangle_L$. Notice that some of these coefficients should be zero. Thus, we need to check the following two things: a) for $l\leqslant \lfloor\frac{L}{2}\rfloor$, where $\lfloor\frac{L}{2}\rfloor$ denotes the greatest integer less or equal to $\frac{L}{2}$,
\begin{equation}
    \sum_{j=0}^{\min\{l,j^{(L)}_{\max}\}} c^{(L)}_j \frac{B_{2l-2j}}{(2l-2j)!} = \frac{(2L-2l)!}{(L-2l)!(2l)!}\,,
\end{equation}
and b) for $l> \lfloor\frac{L}{2}\rfloor$,
\begin{equation}
    \sum_{j=0}^{j^{(L)}_{\max}}c^{(L)}_j \frac{B_{2l-2j}}{(2l-2j)!} = 0\,.
\end{equation}
The above two conditions can be combined into to the following equation:
\begin{equation}
    \sum_{k=0}^{\min\{2l+1,L\}} \frac{(2L-k)!}{k!(L-k)!}\frac{B_{2l+1-k}}{(2l+1-k)!}=0\,.\label{eq:identitytobeproved}
\end{equation}
We need to check this for all $0\leqslant l<L$. Consider the following series
\begin{equation}
    \sum_{n=0}^\infty a_n t^n = \left(\sum_{k=0}^L\frac{(2L-k)!}{k!(L-k)!}t^k\right) \left(\sum_{m=0}^\infty \frac{B_m}{m!}t^m\right)\,,
\end{equation}
then for $l<L$, $a_{2l+1}$ is precisely the l.h.s. of \eqref{eq:identitytobeproved}. Using the definitions of Bessel polynomials (see (\ref{BesselPol}) in Appendix \ref{app:cumulants}) and of the Bernoulli numbers, we have
\begin{equation}
    \sum_{k=0}^L\frac{(2L-k)!}{k!(L-k)!}t^k = \frac{1}{\sqrt{\pi}}e^{\frac{t}{2}}t^{L+\frac{1}{2}}K_{L+\frac{1}{2}}\left(\frac{t}{2}\right)\,,
\end{equation}
and
\begin{equation}
    \sum_{m=0}^\infty \frac{B_m}{m!}t^m = \frac{t}{e^t-1}\,.
\end{equation}
Thus,
\begin{equation}
    \sqrt{\pi}\sum_{n=0}^\infty a_n t^n = \frac{t}{2\sinh\frac{t}{2}} t^{L+\frac{1}{2}}K_{L+\frac{1}{2}}\left(\frac{t}{2}\right) \,.
\end{equation}
The factor $t/2\sinh(t/2)$ is an even function in $t$. The remaining factor can be expanded as follows:
\begin{eqnarray}
    t^{L+\frac{1}{2}}K_{L+\frac{1}{2}}\left(\frac{t}{2}\right) &=& t^{L+\frac{1}{2}} \frac{(-)^L\pi}{2}\left[I_{-L-\frac{1}{2}}\left(\frac{t}{2}\right)-I_{L+\frac{1}{2}}\left(\frac{t}{2}\right)\right]\,,
\end{eqnarray}    
and it can be seen that it contains odd powers of $t$ only starting from the order $t^{2L+1}$. This means that for $l<L$, $a_{2l+1}=0$, and thus \eqref{eq:identitytobeproved} is an identity. So we have proved that
\begin{equation}
    \langle \cQ(z,\bz) \rangle_L = 2i\sum_{j=0}^{\lceil\frac{L}{2}\rceil-1} \frac{2(2L-2j-1)!}{(L-2j-1)!(2j+1)!} (2\ln|z|)^{2j} \mathcal{L}_{2L-2j}(z)\,.
\end{equation}
Using similar calculations we can also show that for $L\geqslant1$ the finite temperature part of $\ln\cZ_L(z,\bz)$, namely the second term in the r.h.s. of (\ref{ZLresult}), can be expanded as
\begin{equation}
    \ln \cZ_L(z,\bz) = 2 \left[\sum_{j=0}^{\lceil\frac{L}{2}\rceil-1} c^{(L)}_j (2\ln|z|)^{2j} \mathcal{L}_{2L+1-2j}(z)-(\sum_{j=0}^{\lceil\frac{L}{2}\rceil-1}c^{(L)}_j \frac{B_{2L-2j}}{(2L-2j)!})(2\ln|z|)^{2L}\mathcal{L}_1 \right]\,.
\end{equation}
On the other hand, for $L=0$, it is straightforward to find
\begin{equation}
    \ln \cZ_0(z,\bz)=Li_1(z)+Li_1(\Bar{z}) =2\mathcal{L}_1\,.
\end{equation}

\section{Two-point functions of massless scalars as thermal cumulants}
\label{app:cumulants}
The recursive applications of $\hat{\,\bf L}_z$ on $\ln \cZ_0$ gives the cumulants of ${\cal Q}$, e.g. the first few ones are
\begin{align}
\label{cumulant0}
\hat{\,\bf L}_z*\ln \cZ_0&=\langle {\cal Q}\rangle_0\,,\\
\hat{\,\bf L}_z^2*\ln \cZ_0&=\langle {\cal Q}^2\rangle_0-\langle{\cal Q}\rangle_0^2\,,\\
\hat{\,\bf L}_z^3*\ln \cZ_0 &= \langle {\cal Q}^3 \rangle_0 - 3\langle {\cal Q} \rangle_0 \langle {\cal Q}^2 \rangle_0 +2 \langle {\cal Q} \rangle_0^3\,.
\end{align}
The quantity $\cD_0^k(z,\bz)$ defined in (\ref{D0k}) can be obtained as a linear combinations of the cumulants of the charge operator ${\cal Q}$, as
\be
\label{cumulant_exp}
\cD_0^k(z,\bz)=\sum_{n=0}^{k-1} \frac{(-)^n(k-1+n)!}{2^n n!(k-1-n)!}\left(\frac{z+\bar{z}}{z-\bar{z}}\right)^n \hat{\,\bf L}_z^{k-n}*\ln Z_0\,.
\ee
The interested reader can verify that the above formula is a reincarnation of the result
\be
\label{invBessels1}
\sin^k\theta\left(\frac{1}{\sin\theta}\frac{\partial}{\partial\theta}\right)^k=\sum_{n=0}^{k-1}{\bf y}^{(k-1)}_n(\cot\theta)\left(\frac{\partial}{\partial\theta}\right)^{k-n}\,,\,\,\,k=1,2,3,..\,,
\ee
with
\be
{\bf y}^{(k)}_n(x)= \frac{(k+n)!}{n!(k-n)!}\frac{x^{n}}{2^n}\,.
\ee
These quantities are the coefficients of the Bessel polynomials defined as
\be
\label{BesselPol}
y_{k}(x)=\sum_{n=0}^{k}{\bf y}_n^{(k)}(x)=\sqrt{\frac{2}{\pi x}}e^{1/x}K_{k+\frac{1}{2}}(1/x)\,.
\ee 
with $K_n(x)$ the modified Bessel functions of the 2nd kind. Perhaps a more transparent form of the result (\ref{cumulant_exp}) is 
\be
\sin^k\theta\left(\frac{1}{\sin\theta}\frac{\partial}{\partial\theta}\right)^k\ln(1+r^2-2r\cos\theta)=\frac{\Gamma(k)}{(1+r^2-2r\cos\theta)^k}\,,
\ee
with $z = re^{i\theta}$.

\section{Proof of the recursive relation}
\label{App:D}
 We prove (\ref{recursion}) by directly studying the thermal one-point functions of the symmetric traceless higher-spin conserved currents $\cO_s$ that appear in the r.h.s. of the thermal two-point function (\ref{GLgL}). Since we are dealing with a free theory, these currents can be readily constructed as
\begin{equation}
    \cO^s_{\mu(s)}=\phi^\dagger(\partial-iA)_{\mu_1}(\partial-iA)_{\mu_2} \cdot\cdot\cdot(\partial-iA)_{\mu_s}\phi + (-)^s h.c.-\text{traces}  \,,\label{eq:currentwithmu}
\end{equation}
where $A_\mu=(\mu,\Vec{0})$ and $\mu(s)$ is a completely symmetric and traceless list of $s$ indices. The thermal one-point functions of such currents in $d$ dimensions\footnote{We use $d$ in this section for convenience with the formulae.} are 
\begin{equation}
    \langle \cO^s_{\mu(s)}\rangle_d = \frac{b_{\cO_s}^{(d)}}{\beta^{\Delta_{\cO_s}}}(e_{\mu_1}e_{\mu_2}\cdot\cdot\cdot e_{\mu_s}-\text{traces})\,,\label{eq:Oonepoint}
\end{equation}
where $e_{\mu}$ is the unit vector in the $\tau$ direction. Hence, it suffices to focus on the $\mu_1=\mu_2=\cdot\cdot\cdot=\mu_s=0$ component. The dimensionless quantity $b_{\cO_s}^{(d)}$ is related to the $a_{\cO_s}^d$ defined in  (\ref{phiphi}) by \cite{Iliesiu:2018fao,Petkou:2018ynm}
\begin{equation}
    a_{\cO_s}^d = \frac{f^{(d)}_{\phi\phi \cO_s}b^{(d)}_{\cO_s}}{c^{(d)}_{\cO_s}}\frac{s!\Gamma(\frac{d-2}{2})}{2^s\Gamma(\frac{d-2}{2}+s)}\,,
\end{equation}
where $c^{(d)}_{\cO_s}$ and $f^{(d)}_{\phi\phi \cO_s}$ are the corresponding two-point and three-point function coefficients. Using $f^{(d)}_{\phi\phi\cO_s}/c^{(d)}_{\cO_s} = 1/s!$, (see e.g. Appendix A of \cite{Benedetti:2023pbt}),  we have
\begin{equation}
    a_{\cO_s}^d = b^{(d)}_{\cO_s}\frac{\Gamma(\frac{d-2}{2})}{2^s\Gamma(\frac{d-2}{2}+s)}\,,
\end{equation}
and the recursive relation \eqref{recursion} becomes
\begin{equation}
    b_{\cO_{s+2}}^{(d)} = 2\pi (d+2s) b_{\cO_s}^{(d+2)} + (\beta m)^2 b_{\cO_s}^{(d)}\,.\label{brecursion}
\end{equation}

To construct the conserved currents (\ref{eq:currentwithmu}) we can use \cite{JARIC20032123} and in particular eq. (3.26) to obtain 
\be
\label{O00s}
\langle \cO^s_{00\cdot\cdot\cdot0} \rangle_d =\langle\phi^\dagger(\partial_0-i\mu)^s\phi\rangle_d +\sum_{n=1}^{\left\lfloor\frac{s}{2}\right\rfloor} \,C_n(d,s)\langle \phi^\dagger (\partial_0-i\mu)^{s-2n}[(\partial-iA)^2]^n\phi\rangle_d + (-)^s c.c.\,,
\ee
where the coefficients $C_n(d,s)$ are
\be
\label{Cs}
C_n(d,s)=\frac{(-1)^n}{\prod_{l=1}^{n}(d+2s-2n+2l-4)} \frac{s!}{(s-2n)!n!2^n}\,,
\ee
and $\lfloor s/2\rfloor$ denotes the greatest integer less or equal to $s/2$. This allows us to use the free equations of motion 
\begin{equation}
    (\partial-iA)^2\phi = m^2\phi\,,
\end{equation}
to obtain
\begin{equation}
\langle \cO^s_{00\cdot\cdot\cdot0} \rangle_d=    \langle \phi^\dagger(\partial_0-i\mu)^s\phi \rangle_d +\sum_{n=1}^{\left\lfloor\frac{s}{2}\right\rfloor} m^{2n}C_n(d,s)\langle \phi^\dagger (\partial_0-i\mu)^{s-2n}\phi \rangle_d + (-)^s c.c.\,.
\end{equation}
Next we notice that performing a Fourier transformation on the spatial coordinates we have for $s\geqslant 2$,
\begin{equation}
\label{ppp}
    \langle\phi^\dagger(-\Vec{\partial}^{\,2})(\partial_0-i\mu)^{s-2}\phi\rangle_d = \int\frac{d^{d-1}\Vec{p}}{(2\pi)^{d-1}}\Vec{p}^{\,2}\langle\phi^\dagger(\tau=0,\Vec{p})(\partial_0-i\mu)^{s-2}\phi(0,-\Vec{p})\rangle\,,
\end{equation}
where
\begin{eqnarray}
    \langle\phi^\dagger(\tau,\Vec{p})(\partial_0-i\mu)^{s-2}\phi(0,-\Vec{p})\rangle &=& \int d^{d-1}\Vec{x}e^{i\Vec{p}\cdot\Vec{x}} \langle\phi^\dagger(\tau,\Vec{x})(\partial_0-i\mu)^{s-2}\phi(0,\Vec{0})\rangle\nonumber\\
    &=& \frac{1}{\beta}\sum_{n=-\infty}^{n=+\infty}\frac{(i\omega-i\mu)^{s-2}e^{-i\omega_n\tau}}{\omega_n^2+\Vec{p}^2+m^2}\,.
\end{eqnarray}
This shows that the integral in (\ref{ppp}) depends only on $|\Vec{p}|$ and we can do the angular integrations to find
\begin{equation}
\label{ppp1}
    \langle\phi^\dagger(-\Vec{\partial}^2)(\partial_0-i\mu)^{s-2}\phi\rangle_d = \frac{S_{d-1}}{(2\pi)^{d-1}}\frac{1}{\beta}\int_0^\infty dp \,p^d \,\sum_{n=-\infty}^{n=+\infty} \frac{(i\omega-i\mu)^{s-2}}{\omega_n^2+p^2+m^2}\,.
\end{equation}
Next, we observe that for $d+2$ dimensions  we have
\begin{align}
\label{ppp2}
 \langle\phi^\dagger(\partial_0-i\mu)^{s-2}\phi\rangle_{d+2}&=   \int\frac{d^{d+1}\Vec{p}}{(2\pi)^{d+1}}\langle\phi^\dagger(\tau=0,\Vec{p})(\partial_0-i\mu)^{s-2}\phi(0,-\Vec{p})\rangle\nonumber \\
 &=\frac{S_{d+1}}{(2\pi)^{d+1}}\frac{1}{\beta}\int_0^\infty dp \,p^d \,\sum_{n=-\infty}^{n=+\infty} \frac{(i\omega-i\mu)^{s-2}}{\omega_n^2+p^2+m^2}\,.
\end{align}
Comparing (\ref{ppp1}) and (\ref{ppp2}) we find
\begin{equation}
\label{ppp3}
    \langle\phi^\dagger(-\Vec{\partial}^2)(\partial_0-i\mu)^{s-2}\phi\rangle_d = 2\pi(d-1)\langle\phi^\dagger(\partial_0-i\mu)^{s-2}\phi\rangle_{d+2}\,.
\end{equation}
Applying the equation of motion in (\ref{ppp3}) we obtain 
\begin{equation}
\label{ppp4}
    2\pi(d-1)\langle\phi^\dagger(\partial_0-i\mu)^{s-2}\phi\rangle_{d+2} = \langle\phi^\dagger(\partial_0-i\mu)^s\phi\rangle_d-m^2\langle\phi^\dagger(\partial_0-i\mu)^{s-2}\phi\rangle_d\,.
\end{equation}
Using (\ref{ppp4}) in the expression of the one-point function $\langle \cO^{s-2}_{00\cdot\cdot\cdot0}\rangle_{d+2}$, we can write it as a linear combination of one-point functions in $d$ dimensions. One can then do a straightforward computation and show the following recursive relation
\begin{equation}
    \langle \cO^s_{00\cdot\cdot\cdot0} \rangle_d = 2\pi (d-1) \langle \cO^{s-2}_{00\cdot\cdot\cdot0}\rangle_{d+2} + m^2\frac{(d+s-4)(d+s-3)}{(d+2s-4)(d+2s-6)}\langle \cO^{s-2}_{00\cdot\cdot\cdot0}\rangle_d\,.\label{Orecursion}
\end{equation}
On the other hand, from \eqref{eq:Oonepoint}, we know that
\begin{eqnarray}
    \langle \cO^s_{00\cdot\cdot\cdot0} \rangle_d &=& \frac{b_{\cO_s}^{(d)}}{\beta^{d+s-2}} (1+\sum_{n=1}^{\left\lfloor\frac{s}{2}\right\rfloor} \frac{(-)^n}{\prod_{l=1}^{n}(d+2s-2n+2l-4)} \frac{s!}{(s-2n)!n!2^n})\nonumber\\
    &=& \frac{b_{\cO_s}^{(d)}}{\beta^{d+s-2}}\prod_{l=1}^{\left\lfloor\frac{s}{2}\right\rfloor}\frac{1}{d+2s-2-2l} \sum_{n=0}^{\left\lfloor\frac{s}{2}\right\rfloor} (-)^n\frac{s!\prod_{k=n+1}^{\left\lfloor\frac{s}{2}\right\rfloor}(d+2s-2-2k)}{(s-2n)!n!2^n}\,.
\end{eqnarray}
The sum over $n$ in the second line can be computed using Mathematica and is found equal to
\begin{equation}
    \frac{2^{\left\lfloor\frac{s}{2}\right\rfloor}\Gamma\left(\frac{d-1}{2}+\left\lfloor\frac{s}{2}\right\rfloor\right)}{\Gamma\left(\frac{d-1}{2}\right)}\,.
\end{equation}
Therefore,
\begin{equation}
    \langle \cO^s_{00\cdot\cdot\cdot0} \rangle_d = \frac{b_{\cO_s}^{(d)}}{\beta^{d+s-2}}\prod_{l=1}^{\left\lfloor\frac{s}{2}\right\rfloor}\frac{d+2\left\lfloor\frac{s}{2}\right\rfloor-1-2l}{d+2s-2-2l}\,.
\end{equation}
Plugging this into  \eqref{Orecursion}  we find \eqref{brecursion}. This derivation works for all $d\geqslant2$ and extends the validity of the result in  \cite{Karydas:2023ufs}. It would be interesting to extend our calculation to non-Gaussian field theories whose fields obey non-linear equations of motion.

 \bibliographystyle{JHEP}
 \bibliography{Refs.bib}

\providecommand{\href}[2]{#2}\begingroup\raggedright\begin{thebibliography}{10}

\bibitem{Symanzik:1972wj}
K.~Symanzik, \emph{{On Calculations in conformal invariant field theories}},
  \href{https://doi.org/10.1007/BF02824349}{\emph{Lett. Nuovo Cim.} {\bfseries
  3} (1972) 734}.

\bibitem{Petkou:1994ad}
A.~Petkou, \emph{{Conserved currents, consistency relations and operator
  product expansions in the conformally invariant O(N) vector model}},
  \href{https://doi.org/10.1006/aphy.1996.0068}{\emph{Annals Phys.} {\bfseries
  249} (1996) 180} [\href{https://arxiv.org/abs/hep-th/9410093}{{\ttfamily
  hep-th/9410093}}].

\bibitem{Hoffmann:2000mx}
L.~Hoffmann, A.C.~Petkou and W.~Ruhl, \emph{{Aspects of the conformal operator
  product expansion in AdS / CFT correspondence}}, {\emph{Adv. Theor. Math.
  Phys.} {\bfseries 4} (2002) 571}
  [\href{https://arxiv.org/abs/hep-th/0002154}{{\ttfamily hep-th/0002154}}].

\bibitem{Hoffmann:2000tr}
L.~Hoffmann, A.C.~Petkou and W.~Ruhl, \emph{{A Note on the analyticity of AdS
  scalar exchange graphs in the crossed channel}},
  \href{https://doi.org/10.1016/S0370-2693(00)00283-5}{\emph{Phys. Lett.}
  {\bfseries B478} (2000) 320}
  [\href{https://arxiv.org/abs/hep-th/0002025}{{\ttfamily hep-th/0002025}}].

\bibitem{Dolan:2000ut}
F.A.~Dolan and H.~Osborn, \emph{{Conformal four point functions and the
  operator product expansion}},
  \href{https://doi.org/10.1016/S0550-3213(01)00013-X}{\emph{Nucl. Phys.}
  {\bfseries B599} (2001) 459}
  [\href{https://arxiv.org/abs/hep-th/0011040}{{\ttfamily hep-th/0011040}}].

\bibitem{Alkalaev:2025zhg}
K.B.~Alkalaev and S.~Mandrygin, \emph{{Multipoint conformal integrals in $D$
  dimensions. Part II: Polygons and basis functions}},
  \href{https://arxiv.org/abs/2507.01904}{{\ttfamily 2507.01904}}.

\bibitem{Alkalaev:2025fgn}
K.B.~Alkalaev and S.~Mandrygin, \emph{{Multipoint conformal integrals in $D$
  dimensions. Part I: Bipartite Mellin-Barnes representation and
  reconstruction}},  \href{https://arxiv.org/abs/2502.12127}{{\ttfamily
  2502.12127}}.

\bibitem{Usyukina:1992jd}
N.I.~Usyukina and A.I.~Davydychev, \emph{{An Approach to the evaluation of
  three and four point ladder diagrams}},
  \href{https://doi.org/10.1016/0370-2693(93)91834-A}{\emph{Phys. Lett. B}
  {\bfseries 298} (1993) 363}.

\bibitem{Usyukina:1993ch}
N.I.~Usyukina and A.I.~Davydychev, \emph{{Exact results for three and four
  point ladder diagrams with an arbitrary number of rungs}},
  \href{https://doi.org/10.1016/0370-2693(93)91118-7}{\emph{Phys. Lett. B}
  {\bfseries 305} (1993) 136}.

\bibitem{Broadhurst:1993ib}
D.J.~Broadhurst, \emph{{Summation of an infinite series of ladder diagrams}},
  \href{https://doi.org/10.1016/0370-2693(93)90202-S}{\emph{Phys. Lett. B}
  {\bfseries 307} (1993) 132}.

\bibitem{Broadhurst:2010ds}
D.J.~Broadhurst and A.I.~Davydychev, \emph{{Exponential suppression with four
  legs and an infinity of loops}},
  \href{https://doi.org/10.1016/j.nuclphysbps.2010.09.014}{\emph{Nucl. Phys. B
  Proc. Suppl.} {\bfseries 205-206} (2010) 326}
  [\href{https://arxiv.org/abs/1007.0237}{{\ttfamily 1007.0237}}].

\bibitem{Abreu:2022mfk}
S.~Abreu, R.~Britto and C.~Duhr, \emph{{The SAGEX review on scattering
  amplitudes Chapter 3: Mathematical structures in Feynman integrals}},
  \href{https://doi.org/10.1088/1751-8121/ac87de}{\emph{J. Phys. A} {\bfseries
  55} (2022) 443004} [\href{https://arxiv.org/abs/2203.13014}{{\ttfamily
  2203.13014}}].

\bibitem{Brown:2004ugm}
F.C.S.~Brown, \emph{{Polylogarithmes multiples uniformes en une variable}},
  \href{https://doi.org/10.1016/j.crma.2004.02.001}{\emph{Compt. Rend. Math.}
  {\bfseries 338} (2004) 527}.

\bibitem{Schnetz:2013hqa}
O.~Schnetz, \emph{{Graphical functions and single-valued multiple
  polylogarithms}},
  \href{https://doi.org/10.4310/CNTP.2014.v8.n4.a1}{\emph{Commun. Num. Theor.
  Phys.} {\bfseries 08} (2014) 589}
  [\href{https://arxiv.org/abs/1302.6445}{{\ttfamily 1302.6445}}].

\bibitem{Isaev:2003tk}
A.P.~Isaev, \emph{{Multiloop Feynman integrals and conformal quantum
  mechanics}}, \href{https://doi.org/10.1016/S0550-3213(03)00393-6}{\emph{Nucl.
  Phys. B} {\bfseries 662} (2003) 461}
  [\href{https://arxiv.org/abs/hep-th/0303056}{{\ttfamily hep-th/0303056}}].

\bibitem{Derkachov:2021ufp}
S.~Derkachov, G.~Ferrando and E.~Olivucci, \emph{{Mirror channel eigenvectors
  of the d-dimensional fishnets}},
  \href{https://doi.org/10.1007/JHEP12(2021)174}{\emph{JHEP} {\bfseries 12}
  (2021) 174} [\href{https://arxiv.org/abs/2108.12620}{{\ttfamily
  2108.12620}}].

\bibitem{Petkou:2018ynm}
A.C.~Petkou and A.~Stergiou, \emph{{Dynamics of Finite-Temperature Conformal
  Field Theories from Operator Product Expansion Inversion Formulas}},
  \href{https://doi.org/10.1103/PhysRevLett.121.071602}{\emph{Phys. Rev. Lett.}
  {\bfseries 121} (2018) 071602}
  [\href{https://arxiv.org/abs/1806.02340}{{\ttfamily 1806.02340}}].

\bibitem{Iliesiu:2018fao}
L.~Iliesiu, M.~Kologlu, R.~Mahajan, E.~Perlmutter and D.~Simmons-Duffin,
  \emph{{The Conformal Bootstrap at Finite Temperature}},
  \href{https://doi.org/10.1007/JHEP10(2018)070}{\emph{JHEP} {\bfseries 10}
  (2018) 070} [\href{https://arxiv.org/abs/1802.10266}{{\ttfamily
  1802.10266}}].

\bibitem{Petkou:2021zhg}
A.C.~Petkou, \emph{{Thermal one-point functions and single-valued
  polylogarithms}},
  \href{https://doi.org/10.1016/j.physletb.2021.136467}{\emph{Phys. Lett. B}
  {\bfseries 820} (2021) 136467}
  [\href{https://arxiv.org/abs/2105.03530}{{\ttfamily 2105.03530}}].

\bibitem{Karydas:2023ufs}
M.~Karydas, S.~Li, A.C.~Petkou and M.~Vilatte, \emph{{Conformal Graphs as
  Twisted Partition Functions}},
  \href{https://doi.org/10.1103/PhysRevLett.132.231601}{\emph{Phys. Rev. Lett.}
  {\bfseries 132} (2024) 231601}
  [\href{https://arxiv.org/abs/2312.00135}{{\ttfamily 2312.00135}}].

\bibitem{Gromov:2018hut}
N.~Gromov, V.~Kazakov and G.~Korchemsky, \emph{{Exact Correlation Functions in
  Conformal Fishnet Theory}},
  \href{https://doi.org/10.1007/JHEP08(2019)123}{\emph{JHEP} {\bfseries 08}
  (2019) 123} [\href{https://arxiv.org/abs/1808.02688}{{\ttfamily
  1808.02688}}].

\bibitem{Loebbert:2022nfu}
F.~Loebbert, \emph{{Integrability for Feynman integrals}},
  \href{https://doi.org/10.21468/SciPostPhysProc.14.008}{\emph{SciPost Phys.
  Proc.} {\bfseries 14} (2023) 008}
  [\href{https://arxiv.org/abs/2212.09636}{{\ttfamily 2212.09636}}].

\bibitem{Kazakov:2018qbr}
V.~Kazakov and E.~Olivucci, \emph{{Biscalar Integrable Conformal Field Theories
  in Any Dimension}},
  \href{https://doi.org/10.1103/PhysRevLett.121.131601}{\emph{Phys. Rev. Lett.}
  {\bfseries 121} (2018) 131601}
  [\href{https://arxiv.org/abs/1801.09844}{{\ttfamily 1801.09844}}].

\bibitem{Giombi:2020enj}
S.~Giombi and J.~Hyman, \emph{{On the large charge sector in the critical O(N)
  model at large N}},
  \href{https://doi.org/10.1007/JHEP09(2021)184}{\emph{JHEP} {\bfseries 09}
  (2021) 184} [\href{https://arxiv.org/abs/2011.11622}{{\ttfamily
  2011.11622}}].

\bibitem{Giombi:2022gjj}
S.~Giombi, E.~Helfenberger and H.~Khanchandani, \emph{{Long range, large
  charge, large N}}, \href{https://doi.org/10.1007/JHEP01(2023)166}{\emph{JHEP}
  {\bfseries 01} (2023) 166}
  [\href{https://arxiv.org/abs/2205.00500}{{\ttfamily 2205.00500}}].

\bibitem{Caetano:2023zwe}
J.a.~Caetano, S.~Komatsu and Y.~Wang, \emph{{Large charge \textquoteright{}t
  Hooft limit of $ \mathcal{N} $ = 4 super-Yang-Mills}},
  \href{https://doi.org/10.1007/JHEP02(2024)047}{\emph{JHEP} {\bfseries 02}
  (2024) 047} [\href{https://arxiv.org/abs/2306.00929}{{\ttfamily
  2306.00929}}].

\bibitem{Derkachov:2018rot}
S.~Derkachov, V.~Kazakov and E.~Olivucci, \emph{{Basso-Dixon Correlators in
  Two-Dimensional Fishnet CFT}},
  \href{https://doi.org/10.1007/JHEP04(2019)032}{\emph{JHEP} {\bfseries 04}
  (2019) 032} [\href{https://arxiv.org/abs/1811.10623}{{\ttfamily
  1811.10623}}].

\bibitem{Basso:2017jwq}
B.~Basso and L.J.~Dixon, \emph{{Gluing Ladder Feynman Diagrams into Fishnets}},
  \href{https://doi.org/10.1103/PhysRevLett.119.071601}{\emph{Phys. Rev. Lett.}
  {\bfseries 119} (2017) 071601}
  [\href{https://arxiv.org/abs/1705.03545}{{\ttfamily 1705.03545}}].

\bibitem{Basso:2021omx}
B.~Basso, L.J.~Dixon, D.A.~Kosower, A.~Krajenbrink and D.-l.~Zhong,
  \emph{{Fishnet four-point integrals: integrable representations and
  thermodynamic limits}},
  \href{https://doi.org/10.1007/JHEP07(2021)168}{\emph{JHEP} {\bfseries 07}
  (2021) 168} [\href{https://arxiv.org/abs/2105.10514}{{\ttfamily
  2105.10514}}].

\bibitem{Gurdogan:2015csr}
O.~G\"urdogan and V.~Kazakov, \emph{{New Integrable 4D Quantum Field Theories
  from Strongly Deformed Planar $\mathcal N = $ 4 Supersymmetric Yang-Mills
  Theory}}, \href{https://doi.org/10.1103/PhysRevLett.117.201602}{\emph{Phys.
  Rev. Lett.} {\bfseries 117} (2016) 201602}
  [\href{https://arxiv.org/abs/1512.06704}{{\ttfamily 1512.06704}}].

\bibitem{Kazakov:2022dbd}
V.~Kazakov and E.~Olivucci, \emph{{The loom for general fishnet CFTs}},
  \href{https://doi.org/10.1007/JHEP06(2023)041}{\emph{JHEP} {\bfseries 06}
  (2023) 041} [\href{https://arxiv.org/abs/2212.09732}{{\ttfamily
  2212.09732}}].

\bibitem{Duhr:2022pch}
C.~Duhr, A.~Klemm, F.~Loebbert, C.~Nega and F.~Porkert,
  \emph{{Yangian-Invariant Fishnet Integrals in Two Dimensions as Volumes of
  Calabi-Yau Varieties}},
  \href{https://doi.org/10.1103/PhysRevLett.130.041602}{\emph{Phys. Rev. Lett.}
  {\bfseries 130} (2023) 041602}
  [\href{https://arxiv.org/abs/2209.05291}{{\ttfamily 2209.05291}}].

\bibitem{Duhr:2023eld}
C.~Duhr, A.~Klemm, F.~Loebbert, C.~Nega and F.~Porkert, \emph{{The Basso-Dixon
  formula and Calabi-Yau geometry}},
  \href{https://doi.org/10.1007/JHEP03(2024)177}{\emph{JHEP} {\bfseries 03}
  (2024) 177} [\href{https://arxiv.org/abs/2310.08625}{{\ttfamily
  2310.08625}}].

\bibitem{Duhr:2023bku}
C.~Duhr and F.~Porkert, \emph{{Feynman integrals in two dimensions and
  single-valued hypergeometric functions}},
  \href{https://doi.org/10.1007/JHEP02(2024)179}{\emph{JHEP} {\bfseries 02}
  (2024) 179} [\href{https://arxiv.org/abs/2309.12772}{{\ttfamily
  2309.12772}}].

\bibitem{Duhr:2024hjf}
C.~Duhr, A.~Klemm, F.~Loebbert, C.~Nega and F.~Porkert, \emph{{Geometry from
  integrability: multi-leg fishnet integrals in two dimensions}},
  \href{https://doi.org/10.1007/JHEP07(2024)008}{\emph{JHEP} {\bfseries 07}
  (2024) 008} [\href{https://arxiv.org/abs/2402.19034}{{\ttfamily
  2402.19034}}].

\bibitem{Brown:2025cbz}
A.~Brown, F.~Galvagno, A.~Grassi, C.~Iossa and C.~Wen, \emph{{Large charge
  meets semiclassics in $\mathcal{N}$ = 4 super Yang-Mills}},
  \href{https://doi.org/10.1007/JHEP06(2025)223}{\emph{JHEP} {\bfseries 06}
  (2025) 223} [\href{https://arxiv.org/abs/2503.02028}{{\ttfamily
  2503.02028}}].

\bibitem{Daoud:2009uj}
M.~Daoud, M.~El~Bouziani, R.~Houca and A.~Jellal, \emph{{Magnetism of Two
  Coupled Harmonic Oscillators}},
  \href{https://doi.org/10.1088/1742-5468/2010/01/P01012}{\emph{J. Stat. Mech.}
  {\bfseries 1001} (2010) P01012}
  [\href{https://arxiv.org/abs/0911.2837}{{\ttfamily 0911.2837}}].

\bibitem{Dunne:2018hog}
G.V.~Dunne, Y.~Tanizaki and M.~{\"U}nsal, \emph{{Quantum Distillation of
  Hilbert Spaces, Semi-classics and Anomaly Matching}},
  \href{https://doi.org/10.1007/JHEP08(2018)068}{\emph{JHEP} {\bfseries 08}
  (2018) 068} [\href{https://arxiv.org/abs/1803.02430}{{\ttfamily
  1803.02430}}].

\bibitem{Haber:1981fg}
H.E.~Haber and H.A.~Weldon, \emph{{Thermodynamics of an Ultrarelativistic Bose
  Gas}}, \href{https://doi.org/10.1103/PhysRevLett.46.1497}{\emph{Phys. Rev.
  Lett.} {\bfseries 46} (1981) 1497}.

\bibitem{Haber:1981tr}
H.E.~Haber and H.A.~Weldon, \emph{{On the Relativistic Bose-einstein
  Integrals}}, \href{https://doi.org/10.1063/1.525239}{\emph{J. Math. Phys.}
  {\bfseries 23} (1982) 1852}.

\bibitem{Haber:1981ts}
H.E.~Haber and H.A.~Weldon, \emph{{Finite Temperature Symmetry Breaking as
  Bose-Einstein Condensation}},
  \href{https://doi.org/10.1103/PhysRevD.25.502}{\emph{Phys. Rev. D} {\bfseries
  25} (1982) 502}.

\bibitem{Goncharov:1995tdt}
A.B.~Goncharov, \emph{{Polylogarithms in Arithmetic and Geometry}},  in
  \emph{{International Congress of Mathematicians}}, 1995,
  \href{https://doi.org/10.1007/978-3-0348-9078-6_31}{DOI}.

\bibitem{Zagier2}
D.~Zagier, \emph{{The Bloch-Wigner-Ramakrishnan polylogarithm function}},
  {\emph{Math. Ann. 286 (1990)} (1990) 613}.

\bibitem{Goncharov:1996dce}
A.~Goncharov, \emph{{Volumes of hyperbolic manifolds and mixed Tate motives}},
  \href{https://arxiv.org/abs/alg-geom/9601021}{{\ttfamily alg-geom/9601021}}.

\bibitem{Ceplak:2021wzz}
N.~Ceplak, S.~Giusto, M.R.R.~Hughes and R.~Russo, \emph{{Holographic
  correlators with multi-particle states}},
  \href{https://doi.org/10.1007/JHEP09(2021)204}{\emph{JHEP} {\bfseries 09}
  (2021) 204} [\href{https://arxiv.org/abs/2105.04670}{{\ttfamily
  2105.04670}}].

\bibitem{Filothodoros:2016txa}
E.G.~Filothodoros, A.C.~Petkou and N.D.~Vlachos, \emph{{$3d$ fermion-boson map
  with imaginary chemical potential}},
  \href{https://doi.org/10.1103/PhysRevD.95.065029}{\emph{Phys. Rev.}
  {\bfseries D95} (2017) 065029}
  [\href{https://arxiv.org/abs/1608.07795}{{\ttfamily 1608.07795}}].

\bibitem{Filothodoros:2018pdj}
E.G.~Filothodoros, A.C.~Petkou and N.D.~Vlachos, \emph{{The fermion-boson map
  for large $d$}},
  \href{https://doi.org/10.1016/j.nuclphysb.2019.01.015}{\emph{Nucl. Phys. B}
  {\bfseries 941} (2019) 195}
  [\href{https://arxiv.org/abs/1803.05950}{{\ttfamily 1803.05950}}].

\bibitem{Drummond:2010cz}
J.M.~Drummond, J.M.~Henn and J.~Trnka, \emph{{New differential equations for
  on-shell loop integrals}},
  \href{https://doi.org/10.1007/JHEP04(2011)083}{\emph{JHEP} {\bfseries 04}
  (2011) 083} [\href{https://arxiv.org/abs/1010.3679}{{\ttfamily 1010.3679}}].

\bibitem{Drummond:2012bg}
J.M.~Drummond, \emph{{Generalised ladders and single-valued polylogarithms}},
  \href{https://doi.org/10.1007/JHEP02(2013)092}{\emph{JHEP} {\bfseries 02}
  (2013) 092} [\href{https://arxiv.org/abs/1207.3824}{{\ttfamily 1207.3824}}].

\bibitem{Alessio:2021krn}
F.~Alessio, G.~Barnich and M.~Bonte, \emph{{Notes on massless scalar field
  partition functions, modular invariance and Eisenstein series}},
  \href{https://doi.org/10.1007/JHEP12(2021)211}{\emph{JHEP} {\bfseries 12}
  (2021) 211} [\href{https://arxiv.org/abs/2111.03164}{{\ttfamily
  2111.03164}}].

\bibitem{Aggarwal:2024axv}
A.~Aggarwal and G.~Barnich, \emph{{Modular properties of massive scalar
  partition functions}},
  \href{https://doi.org/10.1007/JHEP09(2024)127}{\emph{JHEP} {\bfseries 09}
  (2024) 127} [\href{https://arxiv.org/abs/2407.02707}{{\ttfamily
  2407.02707}}].

\bibitem{Dorigoni:2021ngn}
D.~Dorigoni, A.~Kleinschmidt and O.~Schlotterer, \emph{{Poincar{\'e} series for
  modular graph forms at depth two. Part II. Iterated integrals of cusp
  forms}}, \href{https://doi.org/10.1007/JHEP01(2022)134}{\emph{JHEP}
  {\bfseries 01} (2022) 134}
  [\href{https://arxiv.org/abs/2109.05018}{{\ttfamily 2109.05018}}].

\bibitem{Dorigoni:2021jfr}
D.~Dorigoni, A.~Kleinschmidt and O.~Schlotterer, \emph{{Poincar{\'e} series for
  modular graph forms at depth two. Part I. Seeds and Laplace systems}},
  \href{https://doi.org/10.1007/JHEP01(2022)133}{\emph{JHEP} {\bfseries 01}
  (2022) 133} [\href{https://arxiv.org/abs/2109.05017}{{\ttfamily
  2109.05017}}].

\bibitem{Berg:2019jhh}
M.~Berg, K.~Bringmann and T.~Gannon, \emph{{Massive deformations of Maass forms
  and Jacobi forms}},
  \href{https://doi.org/10.4310/CNTP.2021.v15.n3.a4}{\emph{Commun. Num. Theor.
  Phys.} {\bfseries 15} (2021) 575}
  [\href{https://arxiv.org/abs/1910.02745}{{\ttfamily 1910.02745}}].

\bibitem{Hartnoll:2025hly}
S.A.~Hartnoll and M.~Yang, \emph{{The Conformal Primon Gas at the End of
  Time}},  \href{https://arxiv.org/abs/2502.02661}{{\ttfamily 2502.02661}}.

\bibitem{Chamon:2011xk}
C.~Chamon, R.~Jackiw, S.-Y.~Pi and L.~Santos, \emph{{Conformal quantum
  mechanics as the CFT$_1$ dual to AdS$_2$}},
  \href{https://doi.org/10.1016/j.physletb.2011.06.023}{\emph{Phys. Lett. B}
  {\bfseries 701} (2011) 503}
  [\href{https://arxiv.org/abs/1106.0726}{{\ttfamily 1106.0726}}].

\bibitem{Laine:2016hma}
M.~Laine and A.~Vuorinen, \emph{{Basics of Thermal Field Theory}}, vol.~925,
  Springer (2016),
  \href{https://doi.org/10.1007/978-3-319-31933-9}{10.1007/978-3-319-31933-9},
  [\href{https://arxiv.org/abs/1701.01554}{{\ttfamily 1701.01554}}].

\bibitem{Petkou:1998fb}
A.C.~Petkou and N.D.~Vlachos, \emph{{Finite size effects and operator product
  expansions in a CFT for d \ensuremath{>} 2}},
  \href{https://doi.org/10.1016/S0370-2693(98)01530-5}{\emph{Phys. Lett. B}
  {\bfseries 446} (1999) 306}
  [\href{https://arxiv.org/abs/hep-th/9803149}{{\ttfamily hep-th/9803149}}].

\bibitem{Miscioscia:2025pjh}
A.~Miscioscia, \emph{{Thermal effects in conformal field theories}}, Ph.D.
  thesis, Universit{\"a}t Hamburg, 2025.
\newblock \href{https://arxiv.org/abs/2508.02531}{{\ttfamily 2508.02531}}.

\bibitem{Kumar:2025txh}
S.~Kumar, \emph{{Large $N$ Wess-Zumino model at finite temperature and large
  chemical potential in $3d$}},
  \href{https://arxiv.org/abs/2503.17999}{{\ttfamily 2503.17999}}.

\bibitem{Alday:2009aq}
L.F.~Alday, D.~Gaiotto and Y.~Tachikawa, \emph{{Liouville Correlation Functions
  from Four-dimensional Gauge Theories}},
  \href{https://doi.org/10.1007/s11005-010-0369-5}{\emph{Lett. Math. Phys.}
  {\bfseries 91} (2010) 167} [\href{https://arxiv.org/abs/0906.3219}{{\ttfamily
  0906.3219}}].

\bibitem{LeFloch:2020uop}
B.~Le~Floch, \emph{{A slow review of the AGT correspondence}},
  \href{https://doi.org/10.1088/1751-8121/ac5945}{\emph{J. Phys. A} {\bfseries
  55} (2022) 353002} [\href{https://arxiv.org/abs/2006.14025}{{\ttfamily
  2006.14025}}].

\bibitem{Coronado:2018ypq}
F.~Coronado, \emph{{Perturbative four-point functions in planar $ \mathcal{N}=4
  $ SYM from hexagonalization}},
  \href{https://doi.org/10.1007/JHEP01(2019)056}{\emph{JHEP} {\bfseries 01}
  (2019) 056} [\href{https://arxiv.org/abs/1811.00467}{{\ttfamily
  1811.00467}}].

\bibitem{Aprile:2025hlt}
F.~Aprile, S.~Giusto and R.~Russo, \emph{{Four-point correlators with BPS bound
  states in AdS$_3$ and AdS$_5$}},
  \href{https://arxiv.org/abs/2503.02855}{{\ttfamily 2503.02855}}.

\bibitem{Aprile:2024lwy}
F.~Aprile, S.~Giusto and R.~Russo, \emph{{Holographic correlators with BPS
  bound states in $\mathcal{N} = 4$ SYM}},
  \href{https://doi.org/10.1103/PhysRevLett.134.091602}{\emph{Phys. Rev. Lett.}
  {\bfseries 134} (2025) 091602}
  [\href{https://arxiv.org/abs/2409.12911}{{\ttfamily 2409.12911}}].

\bibitem{David:2023uya}
J.R.~David and S.~Kumar, \emph{{Thermal one-point functions:
  CFT\textquoteright{}s with fermions, large d and large spin}},
  \href{https://doi.org/10.1007/JHEP10(2023)143}{\emph{JHEP} {\bfseries 10}
  (2023) 143} [\href{https://arxiv.org/abs/2307.14847}{{\ttfamily
  2307.14847}}].

\bibitem{David:2024naf}
J.R.~David and S.~Kumar, \emph{{One point functions in large N vector models at
  finite chemical potential}},
  \href{https://doi.org/10.1007/JHEP01(2025)080}{\emph{JHEP} {\bfseries 01}
  (2025) 080} [\href{https://arxiv.org/abs/2406.14490}{{\ttfamily
  2406.14490}}].

\bibitem{Diatlyk:2023msc}
O.~Diatlyk, F.K.~Popov and Y.~Wang, \emph{{Beyond $N=\infty$ in Large $N$
  Conformal Vector Models at Finite Temperature}},
  \href{https://arxiv.org/abs/2309.02347}{{\ttfamily 2309.02347}}.

\bibitem{Benedetti:2023pbt}
D.~Benedetti, R.~Gurau, S.~Harribey and D.~Lettera, \emph{{Finite-size versus
  finite-temperature effects in the critical long-range O(N) model}},
  \href{https://doi.org/10.1007/JHEP02(2024)078}{\emph{JHEP} {\bfseries 02}
  (2024) 078} [\href{https://arxiv.org/abs/2311.04607}{{\ttfamily
  2311.04607}}].

\bibitem{David:2024pir}
J.R.~David and S.~Kumar, \emph{{The large N vector model on S$^{1}$
  {\texttimes} S$^{2}$}},
  \href{https://doi.org/10.1007/JHEP03(2025)169}{\emph{JHEP} {\bfseries 03}
  (2025) 169} [\href{https://arxiv.org/abs/2411.18509}{{\ttfamily
  2411.18509}}].

\bibitem{Alkalaev:2024jxh}
K.~Alkalaev and S.~Mandrygin, \emph{{One-point thermal conformal blocks from
  four-point conformal integrals}},
  \href{https://doi.org/10.1007/JHEP10(2024)241}{\emph{JHEP} {\bfseries 10}
  (2024) 241} [\href{https://arxiv.org/abs/2407.01741}{{\ttfamily
  2407.01741}}].

\bibitem{Buric:2024kxo}
I.~Buric, F.~Russo, V.~Schomerus and A.~Vichi, \emph{{Thermal one-point
  functions and their partial wave decomposition}},
  \href{https://doi.org/10.1007/JHEP12(2024)021}{\emph{JHEP} {\bfseries 12}
  (2024) 021} [\href{https://arxiv.org/abs/2408.02747}{{\ttfamily
  2408.02747}}].

\bibitem{Buric:2025uqt}
I.~Buri{\'c}, F.~Mangialardi, F.~Russo, V.~Schomerus and A.~Vichi,
  \emph{{Heavy-Heavy-Light Asymptotics from Thermal Correlators}},
  \href{https://arxiv.org/abs/2506.21671}{{\ttfamily 2506.21671}}.

\bibitem{Gobeil:2018fzy}
Y.~Gobeil, A.~Maloney, G.S.~Ng and J.-q.~Wu, \emph{{Thermal Conformal Blocks}},
  \href{https://doi.org/10.21468/SciPostPhys.7.2.015}{\emph{SciPost Phys.}
  {\bfseries 7} (2019) 015} [\href{https://arxiv.org/abs/1802.10537}{{\ttfamily
  1802.10537}}].

\bibitem{Marchetto:2023xap}
E.~Marchetto, A.~Miscioscia and E.~Pomoni, \emph{{Sum rules {\&} Tauberian
  theorems at finite temperature}},
  \href{https://doi.org/10.1007/JHEP09(2024)044}{\emph{JHEP} {\bfseries 09}
  (2024) 044} [\href{https://arxiv.org/abs/2312.13030}{{\ttfamily
  2312.13030}}].

\bibitem{Barrat:2025wbi}
J.~Barrat, E.~Marchetto, A.~Miscioscia and E.~Pomoni, \emph{{Thermal Bootstrap
  for the Critical O(N) Model}},
  \href{https://doi.org/10.1103/PhysRevLett.134.211604}{\emph{Phys. Rev. Lett.}
  {\bfseries 134} (2025) 211604}
  [\href{https://arxiv.org/abs/2411.00978}{{\ttfamily 2411.00978}}].

\bibitem{Barrat:2025nvu}
J.~Barrat, D.N.~Bozkurt, E.~Marchetto, A.~Miscioscia and E.~Pomoni, \emph{{The
  analytic bootstrap at finite temperature}},
  \href{https://arxiv.org/abs/2506.06422}{{\ttfamily 2506.06422}}.

\bibitem{Buric:2025anb}
I.~Buri{\'c}, I.~Gusev and A.~Parnachev, \emph{{Thermal holographic correlators
  and KMS condition}},  \href{https://arxiv.org/abs/2505.10277}{{\ttfamily
  2505.10277}}.

\bibitem{Shaghoulian:2015kta}
E.~Shaghoulian, \emph{{Modular forms and a generalized Cardy formula in higher
  dimensions}}, \href{https://doi.org/10.1103/PhysRevD.93.126005}{\emph{Phys.
  Rev. D} {\bfseries 93} (2016) 126005}
  [\href{https://arxiv.org/abs/1508.02728}{{\ttfamily 1508.02728}}].

\bibitem{Shaghoulian:2016gol}
E.~Shaghoulian, \emph{{Modular Invariance of Conformal Field Theory on
  $S^1×S^3$ and Circle Fibrations}},
  \href{https://doi.org/10.1103/PhysRevLett.119.131601}{\emph{Phys. Rev. Lett.}
  {\bfseries 119} (2017) 131601}
  [\href{https://arxiv.org/abs/1612.05257}{{\ttfamily 1612.05257}}].

\bibitem{Allameh:2024qqp}
K.~Allameh and E.~Shaghoulian, \emph{{Modular invariance and thermal effective
  field theory in CFT}},
  \href{https://doi.org/10.1007/JHEP01(2025)200}{\emph{JHEP} {\bfseries 01}
  (2025) 200} [\href{https://arxiv.org/abs/2402.13337}{{\ttfamily
  2402.13337}}].

\bibitem{Brust:2016gjy}
C.~Brust and K.~Hinterbichler, \emph{{Free {\ensuremath{\square}}$^{k}$ scalar
  conformal field theory}},
  \href{https://doi.org/10.1007/JHEP02(2017)066}{\emph{JHEP} {\bfseries 02}
  (2017) 066} [\href{https://arxiv.org/abs/1607.07439}{{\ttfamily
  1607.07439}}].

\bibitem{Vanhove:2018elu}
P.~Vanhove and F.~Zerbini, \emph{{Single-valued hyperlogarithms, correlation
  functions and closed string amplitudes}},
  \href{https://doi.org/10.4310/ATMP.2022.v26.n2.a5}{\emph{Adv. Theor. Math.
  Phys.} {\bfseries 26} (2022) 455}
  [\href{https://arxiv.org/abs/1812.03018}{{\ttfamily 1812.03018}}].

\bibitem{Charlton:2021uhu}
S.~Charlton, C.~Duhr and H.~Gangl, \emph{{Clean Single-Valued Polylogarithms}},
  \href{https://doi.org/10.3842/SIGMA.2021.107}{\emph{SIGMA} {\bfseries 17}
  (2021) 107} [\href{https://arxiv.org/abs/2104.04344}{{\ttfamily
  2104.04344}}].

\bibitem{Travaglini:2022uwo}
G.~Travaglini et~al., \emph{{The SAGEX review on scattering amplitudes}},
  \href{https://doi.org/10.1088/1751-8121/ac8380}{\emph{J. Phys. A} {\bfseries
  55} (2022) 443001} [\href{https://arxiv.org/abs/2203.13011}{{\ttfamily
  2203.13011}}].

\bibitem{Loebbert:2024fsj}
F.~Loebbert and S.F.~Stawinski, \emph{{Conformal four-point integrals:
  recursive structure, Toda equations and double copy}},
  \href{https://doi.org/10.1007/JHEP11(2024)092}{\emph{JHEP} {\bfseries 11}
  (2024) 092} [\href{https://arxiv.org/abs/2408.15331}{{\ttfamily
  2408.15331}}].

\bibitem{Dixon:2025zwj}
L.J.~Dixon and C.~Duhr, \emph{{Antipodal self-duality of square fishnet
  graphs}}, \href{https://doi.org/10.1103/PhysRevD.111.L101901}{\emph{Phys.
  Rev. D} {\bfseries 111} (2025) L101901}
  [\href{https://arxiv.org/abs/2502.00862}{{\ttfamily 2502.00862}}].

\bibitem{Caron-Huot:2018dsv}
S.~Caron-Huot, L.J.~Dixon, M.~von Hippel, A.J.~McLeod and G.~Papathanasiou,
  \emph{{The Double Pentaladder Integral to All Orders}},
  \href{https://doi.org/10.1007/JHEP07(2018)170}{\emph{JHEP} {\bfseries 07}
  (2018) 170} [\href{https://arxiv.org/abs/1806.01361}{{\ttfamily
  1806.01361}}].

\bibitem{He:2025lzd}
S.~He and X.~Jiang, \emph{{Solving Infinite Families of Dual Conformal
  Integrals and Periods}},  \href{https://arxiv.org/abs/2506.20095}{{\ttfamily
  2506.20095}}.

\bibitem{DeClerck:2025mem}
M.~De~Clerck, S.A.~Hartnoll and M.~Yang, \emph{{Wheeler-DeWitt wavefunctions
  for 5d BKL dynamics, automorphic L-functions and complex primon gases}},
  \href{https://arxiv.org/abs/2507.08788}{{\ttfamily 2507.08788}}.

\bibitem{Olivucci:2021cfy}
E.~Olivucci, \emph{{Hexagonalization of Fishnet integrals. Part I. Mirror
  excitations}}, \href{https://doi.org/10.1007/JHEP11(2021)204}{\emph{JHEP}
  {\bfseries 11} (2021) 204}
  [\href{https://arxiv.org/abs/2107.13035}{{\ttfamily 2107.13035}}].

\bibitem{Derkachov:2023xqq}
S.E.~Derkachov, A.P.~Isaev and L.A.~Shumilov, \emph{{Ladder and zig-zag Feynman
  diagrams, operator formalism and conformal triangles}},
  \href{https://doi.org/10.1007/JHEP06(2023)059}{\emph{JHEP} {\bfseries 06}
  (2023) 059} [\href{https://arxiv.org/abs/2302.11238}{{\ttfamily
  2302.11238}}].

\bibitem{Gradshteyn:1943cpj}
I.S.~Gradshteyn and I.M.~Ryzhik, \emph{{Table of Integrals, Series, and
  Products}}, Academic Press (2007 7th ed.).

\bibitem{JARIC20032123}
J.~Jari{\'c}, \emph{On the decomposition of symmetric tensors into traceless
  symmetric tensors},
  \href{https://doi.org/https://doi.org/10.1016/S0020-7225(03)00202-7}{\emph{International
  Journal of Engineering Science} {\bfseries 41} (2003) 2123}.

\end{thebibliography}\endgroup

\end{document}